\documentclass[aps,twocolumn,prl]{revtex4-2}

\usepackage{xcolor,hyperref}
\hypersetup{
   colorlinks,
   linkcolor={blue!50!black},
   citecolor={blue!50!black},
   urlcolor={blue!80!black}
}

\usepackage{epsfig, amsmath}
\usepackage{bm}
\usepackage{soul}
\usepackage{hyperref}
\usepackage{footmisc}
\DeclareMathAlphabet{\mathitbf}{OML}{cmm}{b}{it}

\renewcommand{\=}{\!=\!}

\newcommand{\xv}{\mathitbf x}
\newcommand{\calBold}[1]{\mbox{\boldmath${\cal #1}$}}
\newcommand{\dbar}{{\,\mathchar'26\mkern-12mu d}}

\begin{document}

\title{Brittle to ductile transitions in glasses: Roles of soft defects and loading geometry}
\author{David Richard$^{1}$}
\email{d.richard@uva.nl}
\author{Edan Lerner$^{1}$}
\email{Corresponding author: e.lerner@uva.nl}
\author{Eran Bouchbinder$^{2}$}
\email{Corresponding author: eran.bouchbinder@weizmann.ac.il}
\affiliation{$^{1}$Institute of Theoretical Physics, University of Amsterdam, Science Park 904, 1098 XH Amsterdam, the Netherlands\\
$^{2}$Chemical and Biological Physics Department, Weizmann Institute of Science, Rehovot 7610001, Israel}

\begin{abstract}
Understanding the fracture toughness of glasses is of prime importance for science and technology. We study it here using extensive atomistic simulations in which the interaction potential, glass transition cooling rate and loading geometry are systematically varied, mimicking a broad range of experimentally accessible properties. Glasses' nonequilibrium mechanical disorder is quantified through $A_{\rm g}$, the dimensionless prefactor of the universal spectrum of nonphononic excitations, which measures the abundance of soft glassy defects that affect plastic deformability. We show that while a brittle-to-ductile transition might be induced by reducing the cooling rate, leading to a reduction in $A_{\rm g}$, iso-$\!A_{\rm g}$ glasses are either brittle or ductile depending on the degree of Poisson contraction under unconstrained uniaxial tension. Eliminating Poisson contraction using constrained tension reveals that iso-$\!A_{\rm g}$ glasses feature similar toughness, and that varying $A_{\rm g}$ under these conditions results in significant toughness variation. Our results highlight the roles played by both soft defects and loading geometry (which affects the activation of defects) in the toughness of glasses.
\end{abstract}

\maketitle

%
%

\section{Introduction}

Glasses are intrinsically nonequilibrium materials whose properties may vary vastly with their thermomechanical history, e.g.~the cooling rate at which they are formed from a melt~\cite{Harmon_apl_2007}. The nonequilibrium nature of glasses is mesoscopically manifested through a broad range of disordered structures, which qualitatively differ from the corresponding ordered structures featured by their crystalline counterparts. Macroscopically, the nonequilibrium nature of glasses may be manifested in a substantial variability in various material properties, even for the very same composition.

\begin{figure*}[ht!]
  \includegraphics[scale=1]{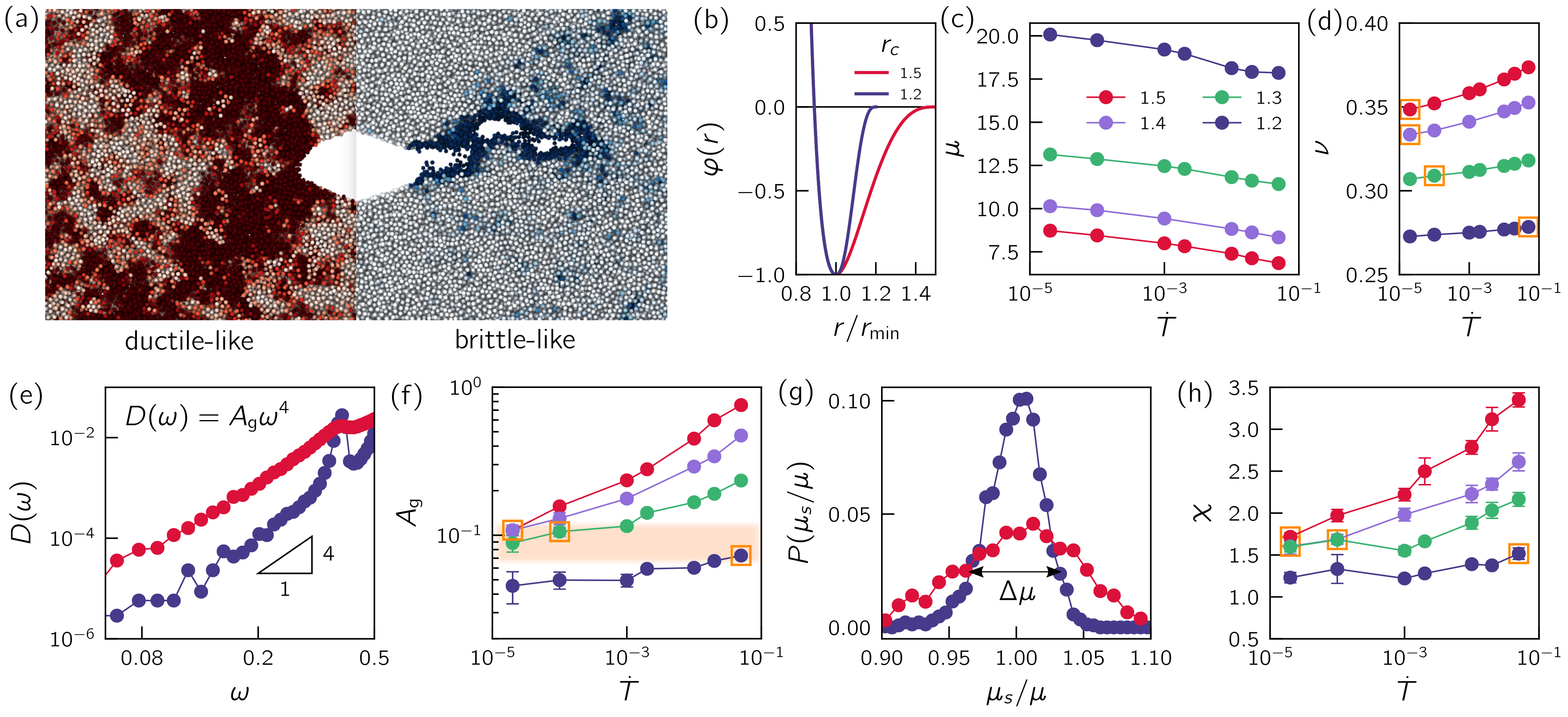}
  \caption{{\bf Computer glasses span a broad range of macroscopic and mesoscopic material properties}. (a) Snapshots of the non-affine plastic deformation during fracture tests with an initial diamond crack for ductile-like (left) and brittle-like (right) failure. The non-affine deformation is quantified by the widely-used $D^2_{\rm min}$ field~\cite{falk_langer_stz}, here and throughout this paper, where a darker color is indicative of more intense plastic deformation. (b) The modified Lennard-Jones type potential $\varphi(r)$ used in this work (see \emph{Methods}). $\varphi$ is plotted vs.~the normalized interparticle distance $r/r_{\rm min}$, where $r_{\rm min}$ is the minimum of the potential. The parameter $r_{\rm c}$ controls the strength and range of the attractive term. (c) The shear modulus $\mu$ (reported in simulational units, see \emph{Methods}) vs.~the quench rate $\dot{T}$ for various $r_{\rm c}$ values. The color code for the latter is detailed in the legend, and will be adopted throughout the paper. (d) Poisson's ratio $\nu$ vs.~$\dot{T}$. The meaning of the open orange squares is explained in panel (f) below. (e) The vibrational density of states (vDOS) of QLMs for $\dot{T}\!=\!0.05$ and two values of $r_{\rm c}$ (cf.~color code in panel (c)). The peak in the vDOS corresponds to the first (phononic) shear wave. (f) The prefactor $A_{\rm g}$ of the universal nonphononic vDOS $D(\omega)\!=\!A_{\rm g}\,\omega^4$, made dimensionless through scaling by $\omega_0^5$, as a function of $\dot{T}$. A narrow band of nearly equal $A_{\rm g}$ values is highlighted by a blurry orange region, and the glasses within this narrow band are marked by open orange squares (the same glasses are marked in panel (d)). (g) The sample-to-sample shear modulus $\mu_{\rm s}$ distributions $P(\mu_{\rm s}/\mu)$ for the same glasses as in (e). $\mu$ is the mean (plotted in panel (c)) and the width of one of the distributions, indicative of the magnitude of shear modulus fluctuations, is marked by $\Delta\mu$. (h) The dimensionless quantifier of shear modulus fluctuations $\chi$ (see \emph{Methods} for exact definition) is plotted against $\dot{T}$ for various $r_{\rm c}$'s. The blurry orange region marks iso-$\chi$ states (similarly to the iso-$\!A_{\rm g}$ states in panel (f)). As stated above, the same color code describing different values of $r_{\rm c}$ is maintained throughout this paper.}
  \label{fig:fig1}
\end{figure*}

One such important material property is the ability of a glass to resist catastrophic failure. The latter is commonly quantified by the fracture toughness, which measures the material resistance to failure in the presence of an initial crack~\cite{lawn1993fracture}. The value of the fracture toughness reflects multitude of spatiotemporal processes that take place in the material prior to catastrophic failure. In one limit, failure is accompanied by rather localized plastic deformation (cf.~Fig.~\ref{fig:fig1}a, right) and is generally quite abrupt. The fracture toughness is typically small in this case, and failure is commonly termed \emph{brittle-like}. In the opposite limit, failure is accompanied by significant spatially-extended plastic deformation (cf.~Fig.~\ref{fig:fig1}a, left) and is generally more gradual. The fracture toughness is typically larger in this case, and failure is commonly termed \emph{ductile-like}.

Understanding the transition between these two modes of failure is an important challenge in materials science~\cite{brittleness_BMG,Greer_2012_AM,shi_2014,Rycroft2012,Vasoya2016,Eran_mechanical_glass_transition}.
Establishing, understanding and predicting structure-dynamics-properties relations in glasses, e.g.~in the context of material failure, involve various challenges. First, one should develop tools to quantify the disordered structures featured by glasses. Second, one should identify a certain sub-class of structural degrees of freedom that is relevant for a given macroscopic material property. Third, one should understand the dynamic evolution of these relevant degrees of freedom under some prescribed conditions. Finally, one should be able to coarse-grain over mesoscopic structural degrees of freedom in order to understand how collective dynamics control the macroscopic material property.

Quantifying mesoscopic structural and mechanical disorder in glasses is challenging. One aspect of the challenge is that we generally lack tools and concepts to distinguish one disordered state from another, in sharp contrast to well established ``order parameters'' in other condensed-matter systems~\cite{sethna2006statistical}. Another aspect of the challenge is the associated length scales; glassy disorder typically manifests itself on small length scales, usually corresponding to a few atomic distances, which are not directly accessible to currently available experimental techniques in molecular glasses. Consequently, computer glasses --- where such length scales are readily accessible --- have played important roles in our understanding of glassy disorder~\cite{Coslovich_2007_lfs,Tanaka_Tong_prl_2020,paddy_huge_review_2015,LB_modes_2019}.

Recently, computer studies revealed and substantiated the existence of low-frequency (soft) nonphononic vibrational modes in glasses~\cite{Schober_Laird_numerics_PRL,Schober_Oligschleger_numerics_PRB,falk_pnas_2014,modes_prl_2016,ikeda_pnas,pinching_pnas,LB_modes_2019,pseudo_harmonic_modes_prl_2021}. These soft vibrational modes (or excitations) have been shown to be quasilocalized in space~\cite{Schober_Laird_numerics_PRL,Schober_Oligschleger_numerics_PRB,core_properties}, as opposed to the spatially-extended nature of low-frequency phonons (i.e.~plane waves), hence they are termed hereafter quasilocalized modes (QLMs). Moreover, QLMs have been shown to follow a universal vibrational density of states (vDOS) ${\cal D}(\omega)\!\sim\!\omega^4$, where $\omega$ is the angular vibrational frequency~\cite{Gurevich2003,Gurevich2007,modes_prl_2016,ikeda_pnas,modes_prl_2018,pinching_pnas,LB_modes_2019,modes_prl_2020,KHGPS_scaling_theory,KHGPS_exact_results} (cf.~examples in Fig.~\ref{fig:fig1}e). As the $\omega^4$ law is universal --- i.e.~independent of the glass composition, interatomic interactions and dimensionality --- all of the glass history dependence is encapsulated in the non-universal prefactor $A_{\rm g}$, defined through ${\cal D}(\omega)\!\equiv\!A_{\rm g}\,\omega^4$. $A_{\rm g}$, once properly non-dimensionalized (see below), has been shown to provide a measure of the number density of QLMs in a glass~\cite{pinching_pnas,LB_modes_2019,mw_thermal_origin,modes_prl_2020,sticky_spheres_part_1}.

QLMs, in turn, have been shown to correlate with the spatial loci of irreversible rearrangements in glasses, i.e.~they statistically correlate with Shear-Transformation-Zones (STZs), the carriers of plasticity in glassy materials~\cite{lemaitre2004,tanguy2010vibrational,manning2011,falk_pnas_2014,rottler_normal_modes,plastic_modes_prerc,david_huge_collaboration,pseudo_harmonic_modes_prl_2021}. As such, the properly non-dimensionalized $A_{\rm g}$ is directly relevant for plastic deformability, and consequently to the fracture toughness, as it provides a measure of the number of soft defects embedded inside glassy structures~\cite{pinching_pnas,LB_modes_2019,mw_thermal_origin,sticky_spheres_part_1}. Another measure of mesocopic mechanical disorder in glasses can be defined using the fluctuations of the shear modulus $\mu$~\cite{tsamados2009local,mizuno2013measuring,scattering_jcp_2020,sticky_spheres_part_1,fsp}. This measure, denoted hereafter by $\chi$ (see \emph{Methods} and Fig.~\ref{fig:fig1}g), has been recently shown to control the rate of wave attenuation in computer glasses~\cite{Schirmacher_2006,Schirmacher_prl_2007,scattering_jcp_2020}, i.e.~the amplitude of Rayleigh scattering rates in the low-frequency (long-wavelength) limit. In this paper, we will be using $A_{\rm g}$ and $\chi$ as quantifiers of mesoscopic glassy disorder, which are directly relevant to the fracture toughness.

$A_{\rm g}$ and $\chi$ allow one to compare on equal footing different glasses, either glasses of the same interatomic interactions/composition formed by different thermal protocols or glasses of different interatomic interactions/composition. While existing evidence clearly indicates that $A_{\rm g}$ and $\chi$ must play important roles in determining the fracture toughness of these materials, one obviously cannot exclude the possibility that other physical quantities and factors play a role as well. Indeed, some experimental and computational studies~\cite{schroers2004ductile,brittleness_BMG,castellero2007critical,greaves2011poisson,Greer_2012_AM,shi_2014,deng2018measuring} suggested that Poisson's ratio $\nu$ plays an important role in determining the fracture toughness of glasses (though some other studies challenged this view~\cite{Eran_Chris_prl_2012,sticky_spheres_part_1,kumar2013critical}). A possible interpretation is that $\nu$ is sensitive to the structure of a glass, and in particular that it may be indirectly related to plastic deformability and its sensitivity to shear versus tensile deformation~\cite{super_plastic_nature_2007,shi_2014}, and hence affects the toughness. As such, one can speculate that the observed effect of $\nu$ is not qualitatively different from that of $A_{\rm g}$ and $\chi$ --- indeed we show below that well-defined relations between $\nu$ and $A_{\rm g}$ (and $\chi$) exist. It remains unclear, however, whether $\nu$ --- or more precisely Poisson contraction/effect and other loading geometry effects --- also play distinct roles in determining the toughness.

Our goal in this paper is to understand the roles played by soft defects, as quantified by $A_{\rm g}$ and indirectly by $\chi$, and by the loading geometry employed in mechanical tests on the fracture toughness of glasses, in particular on brittle-to-ductile transitions. This goal is achieved by employing extensive computer glass simulations in 3D, which offer a powerful and flexible platform to address the posed questions. To mimic glasses' compositional variability we employ computer models based on Lennard-Jones interactomic pairwise potentials with a \emph{tunable} parameter $r_{\rm c}$ (to be accurately defined below), shown recently to give rise to glasses of widely variable properties~\cite{itamar_brittle_to_ductile_pre_2011,sticky_spheres_part_1,sticky_spheres_part_2,minimally_disordered_glasses_arXiv}. We also apply a broad range of cooling rates $\dot{T}$ across the glass transition (to be accurately defined below). In addition, we employ cutting-edge algorithms that allow to generate deeply supercooled glasses~\cite{fsp}, to a degree that is comparable or even surpasses that of laboratory glasses. Overall, the range of variability of the macroscopic properties (e.g.~$\mu$ and $\nu$, cf.~Fig.~\ref{fig:fig1}c-d) of our computer glasses is comparable to that observed in laboratory glasses. Finally, we carefully and systematically vary the imposed loading geometry in the fracture toughness tests in order to quantify its effect on the toughness.

We find that the fracture toughness, and in particular brittle-to-ductile transitions, are controlled by \emph{both} the abundance of soft defects --- as quantified by $A_{\rm g}$ (and indirectly by $\chi$) --- and by the loading geometry of the fracture test. The latter affects the relative magnitude of shear and tensile deformation experienced by the material, and consequently the emerging plastic dissipation, for a fixed $A_{\rm g}$. That is, the loading geometry controls the activation of soft defects, whose abundance is controlled by $A_{\rm g}$. Only under a certain choice of loading geometry, where Poisson contraction can take place, Poisson's ratio $\nu$ can be sensibly used to quantify the fracture toughness \emph{together} with $A_{\rm g}$ (or $\chi$). These results provide basic insights into the physical origin of the failure resistance of glasses.

\section{Results}

In order to study the origin of brittle-to-ductile transitions in glasses, such as the one illustrated in Fig.~\ref{fig:fig1}a, we first aim at generating computer glasses featuring a broad range of properties. This is achieved by employing the computer glass model put forward in~\cite{itamar_brittle_to_ductile_pre_2011}, where particles interact via a tunable Lennard-Jones-like pairwise potential (a similar approach was taken in~\cite{falk1999molecular}). In particular, the strength and range of the attractive term of the pairwise potential are controlled by a parameter $r_{\rm c}$, as shown in Fig.~\ref{fig:fig1}b and accurately defined in \emph{Methods}. Varying $r_{\rm c}$, which may qualitatively correspond to varying the glass composition, results in dramatic changes in emergent material properties, as will be discussed and demonstrated soon (see also~\cite{sticky_spheres_part_1,sticky_spheres_part_2}). In addition, we generate different glasses by
quenching equilibrium liquids at a broad range of rates into their arrested glass phase. The cooling rate is quantified by $\dot{T}$, the dimensionless absolute value of the rate of change of temperature during the quench (see \emph{Methods} for precise definition). Yet another glass formation protocol, which is not quantified by $\dot{T}$, is employed and discussed below.

In Fig.~\ref{fig:fig1}c, we present the shear modulus $\mu$ (in simulation units, see \emph{Methods}) as a function of $\dot{T}$ for various values of $r_{\rm c}$. The corresponding results for Poisson's ratio $\nu$ are shown in Fig.~\ref{fig:fig1}d. The range of variation in $\nu$ accessed by tuning $r_{\rm c}$ compares well with the one for which the so-called ductile-to-brittle transition is observed in bulk metallic glasses \cite{wang2014understanding}, and see \emph{Supplementary material} for further comparisons between our model's elastic properties and typical laboratory glasses' elastic properties. While our model's macroscopic linear response coefficients make direct contact with laboratory glasses, computer glasses offer unique access to various physical quantities that are not directly accessible to experiments, as highlighted above. In Fig.~\ref{fig:fig1}e, we present the vDOS of QLMs for two values of $r_{\rm c}$ and a fixed $\dot{T}$, both revealing the universal ${\cal D}(\omega)\=A_{\rm g}\,\omega^4$ law. Here and elsewhere, we report frequencies in terms of $\omega_0\!\equiv\!c_s/a_0$, where $c_s$ is the (zero frequency) shear wave speed and $a_0\!\equiv\!(V/N)^{1/3}$ ($V$ is the system's volume containing $N$ particles). The observed prefactor $A_{\rm g}$ vastly varies between the two cases. $A_{\rm g}$, made dimensionless through multiplication by $\omega_0^5$, is presented in Fig.~\ref{fig:fig1}f for all $\dot{T}$ and $r_{\rm c}$ values used in Fig.~\ref{fig:fig1}c-d. The dimensionless $A_{\rm g}$ (hereafter we exclusively refer to the dimensionless one) indeed reveals large variability. Interestingly, different combinations of $\dot{T}$ and $r_{\rm c}$ result in nearly identical $A_{\rm g}$ values (marked by open orange squares), a property that will be used below.
\begin{figure}[t!]
  \includegraphics[scale=1]{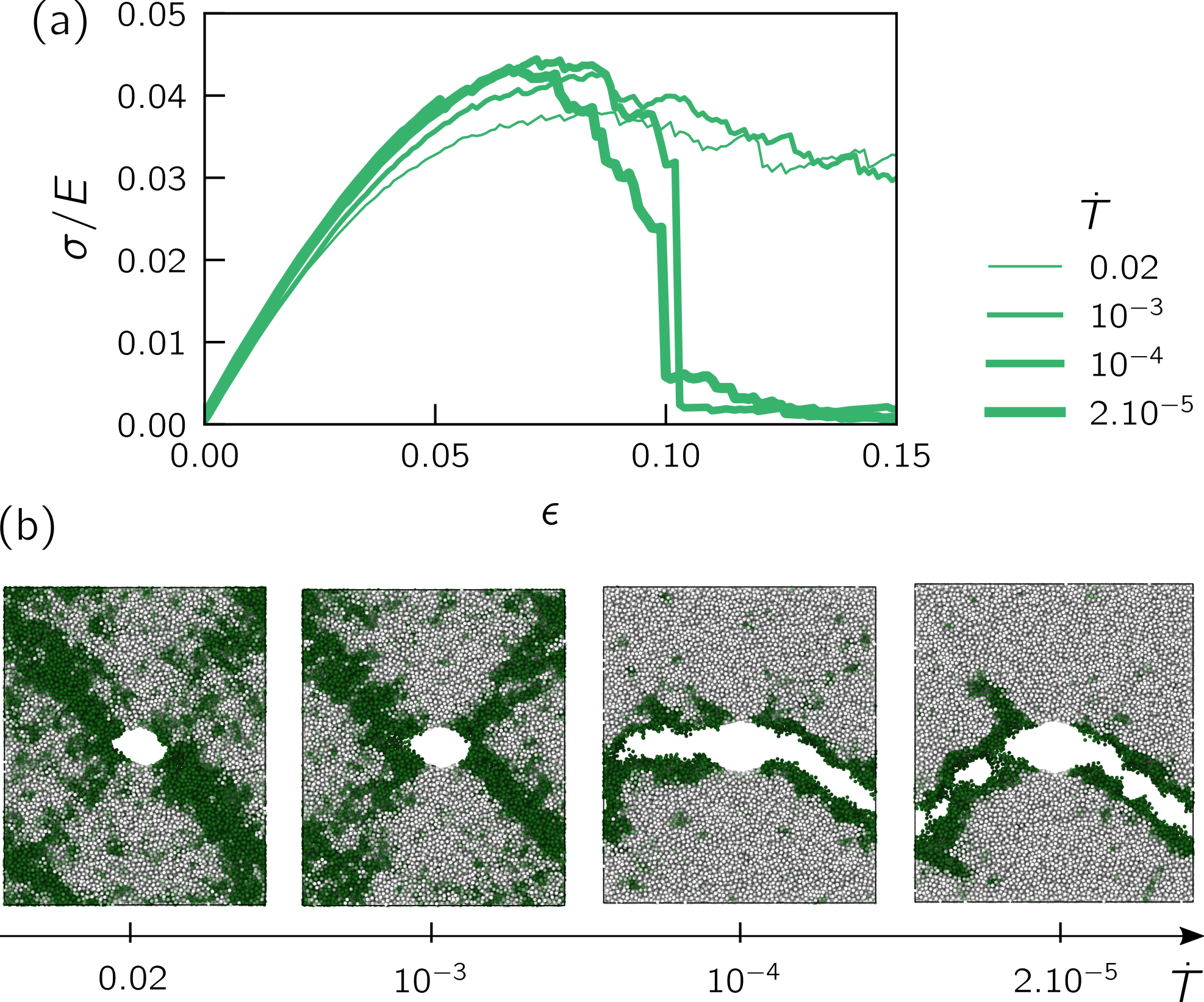}
  \caption{{\bf A ductile-to-brittle transition induced by varying the cooling rate through the glass transition}.  (a) Tensile stress-strain curves (unconstrained uniaxial tension) of glasses with $r_{\rm c}\!=\!1.3$, prepared at various quench rates $\dot{T}$ (see legend). (b) The corresponding non-affine plastic deformation computed at $\epsilon\!=\!0.15$.}
  \label{fig:fig2}
\end{figure}

\begin{figure*}[t!]
  \includegraphics[scale=1]{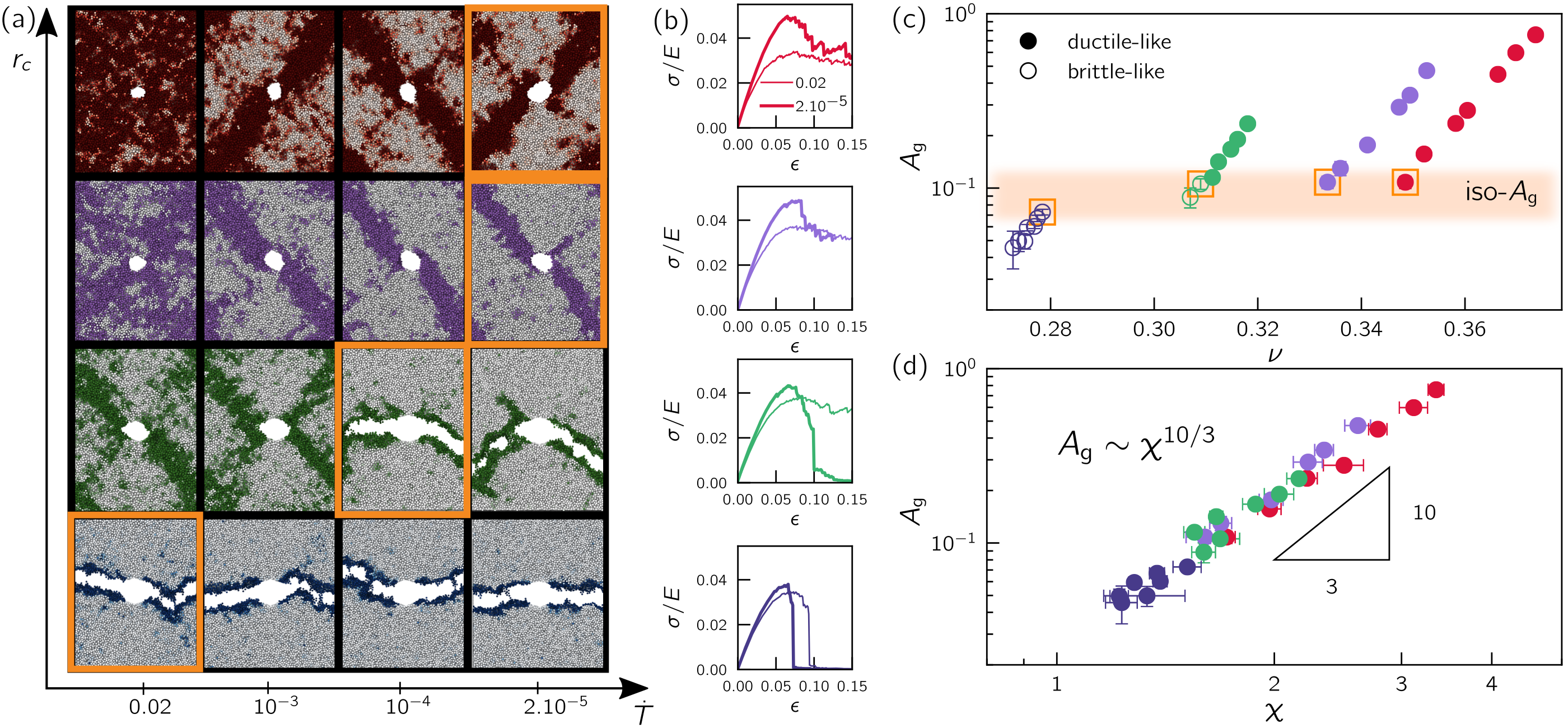}
  \caption{{\bf The fracture toughness depends on both the number of soft defects, quantified by $A_{\rm g}$, and on Poisson contraction, quantified by $\nu$, under unconstrained uniaxial tension}. (a) Snapshots of the non-affine plastic deformation at $15\%$ strain in the $r_{\rm c}\!-\!\dot{T}$ parametric plane. (b) Strain-strain curve for a poorly-annealed (high $\dot{T}$, thin curves) and well-annealed (low $\dot{T}$, thick curves) glasses with various $r_c$. (c) $A_{\rm g}$ vs.~$\nu$, where ductile-like glasses correspond to full symbols and brittle-like to empty ones.  The orange squares and the blurry region indicate glasses with a nearly identical density of OLMs (iso-$\!A_{\rm g}$). (d) $A_{\rm g}$ vs.~$\chi$, roughly following a power-law relation $A_{\rm g}\!\sim\!\chi^{10/3}$ for all glasses (see discussion in the \emph{Supplementary~material}).}
  \label{fig:fig3}
\end{figure*}

$A_{\rm g}$, providing a measure of the abundance of soft defects inside a glass~\cite{pinching_pnas}, serves as a disorder quantifier that is relevant for plastic deformability and fracture toughness. Another quantifier of mechanical disorder in glasses can be obtained by considering the distribution of the local shear modulus, probed by studying a large number of systems of size $N$~\cite{scattering_jcp_2020,minimally_disordered_glasses_arXiv}. In Fig.~\ref{fig:fig1}h we present the shear modulus distributions corresponding to the two case shown in Fig.~\ref{fig:fig1}e. It is observed that smaller $A_{\rm g}$ glasses feature reduced shear modulus fluctuations $\Delta\mu/\mu$. The latter can be used to construct an $N$-independent quantifier of mechanical disorder in the form $\chi\!\equiv\!(\Delta\mu/\mu)\sqrt{N}$~\cite{Schirmacher_2006,Schirmacher_prl_2007,minimally_disordered_glasses_arXiv,scattering_jcp_2020}. The dependence of $\chi$ on both $r_{\rm c}$ and $\dot{T}$ is presented Fig.~\ref{fig:fig1}g, and the relation between $\chi$ and $A_{\rm g}$ is further discussed below (cf.~Fig.~\ref{fig:fig3}d and \emph{Supplementary material}). Next, we start considering the fracture toughness of the glasses at hand.

\subsection{Reduction in cooling rate can induce a ductile-to-brittle transition}

In order to probe the fracture toughness we consider 3D glass samples containing an initial central crack of fixed geometry that cuts through the sample. In particular, we choose a diamond-shaped crack of fixed length and main vertex angle, which give rise to traction-free crack surfaces. The main vertex angle is chosen to be sufficiently small in order to reproduce the classical square root singularity of linear elastic fracture mechanics~\cite{lawn1993fracture}, see \emph{Supplementary material} for additional details. The initial crack length is chosen to be sufficiently larger than the particle size and sufficiently smaller than the sample's dimensions in order to ensure reasonable scale separation, though we cannot entirely exclude any finite size effects (see \emph{Supplementary material}). All fracture toughness calculations are performed under quasi-static athermal conditions, i.e.~vanishing applied strain-rate and zero temperature, as our focus is on the effect of glass structure and loading geometry on the toughness.

We first consider the effect of varying the cooling rate $\dot{T}$ on the glass resistance to failure under unconstrained uniaxial tension, where `unconstrained' means that the lateral edges (perpendicular to the tensile axis) are traction-free and hence free to contract. We focus here on the $r_{\rm c}\=1.3$ glasses ensemble, and plot in Fig.~\ref{fig:fig2}a the tensile stress-strain curves for four different cooling rates $\dot{T}$ (see legend). It is observed that for the two largest $\dot{T}$ values, the corresponding glasses feature gradual stress relaxation after reaching a peak stress, with no clear sign of catastrophic failure. For the two smallest $\dot{T}$ values, however, the glass appears to lose its load-bearing capacity at a well-defined strain level, indicating catastrophic failure. Consequently, there appears to exist a cooling rate that induces a ductile-to-brittle transition, as was also demonstrated in previous work~\cite{yuan2012molecular,li2015correlation,Eran_mechanical_glass_transition,lin2019distinguishing}. This is corroborated by the spatiotemporal dynamics of the deformed samples, as demonstrated in Fig.~\ref{fig:fig2}b, where snapshots at $15\%$ strain are shown. Consistent with the stress-strain curves, we observe that while glasses formed under the two highest cooling rates feature a blunted crack accompanied by large scale plastic deformation (darker regions, see figure caption), glasses cooled at the two lowest rates feature a crack that propagates through the samples, accompanied by rather localized plastic deformation.

It is known that varying the cooling rate $\dot{T}$ leads to varying glass structures and consequently to variation in many material properties of a glass~\cite{eran_cooling_rate_G_PRE_2013,cge_paper,LB_modes_2019}. Our next goal is to understand which physical properties control the ductile-to-brittle transition observed in Fig.~\ref{fig:fig2}, when $\dot{T}$ is varied.

\subsection{Do soft defects exclusively control the ductile-to-brittle transition?
The role of Poisson contraction under unconstrained uniaxial tension}

As explained in the Introduction, $A_{\rm g}$ --- whose $\dot{T}$ dependence is presented in Fig.~\ref{fig:fig1}f --- is a natural candidate for controlling the observed ductile-to-brittle transition. To test this natural expectation, we cannot restrict ourselves to a single $r_{\rm c}$ value; rather, we employ glasses cooled at different quench rates $\dot{T}$ \emph{and} of different interaction parameters $r_{\rm c}$ that nevertheless share nearly the \emph{same} (dimensionless) $A_{\rm g}$. These glasses, which are marked by open orange squares in Fig.~\ref{fig:fig1}f, are expected to feature similar resistance to failure --- if indeed $A_{\rm g}$ fully controls the fracture toughness of glasses.
In Fig.~\ref{fig:fig3}a we plot $15\%$-strain snapshots in the $r_{\rm c}\!-\!\dot{T}$ parametric plane (covering the whole range of employed parameters) --- similarly to Fig.~\ref{fig:fig2}b (in fact, the latter is reproduced here in the third row from the top) ---, where the nearly iso-$\!A_{\rm g}$ are highlighted by open orange rectangles. It is observed that iso-$\!A_{\rm g}$ glasses are either brittle-like or ductile-like, indicating that $\!A_{\rm g}$ does not exclusively control the fracture toughness under the present conditions (i.e.~under unconstrained uniaxial tension). This conclusion is corroborated by the stress-strain curves presented in Fig.~\ref{fig:fig3}b.
\begin{figure*}[t!]
 \includegraphics[scale=1]{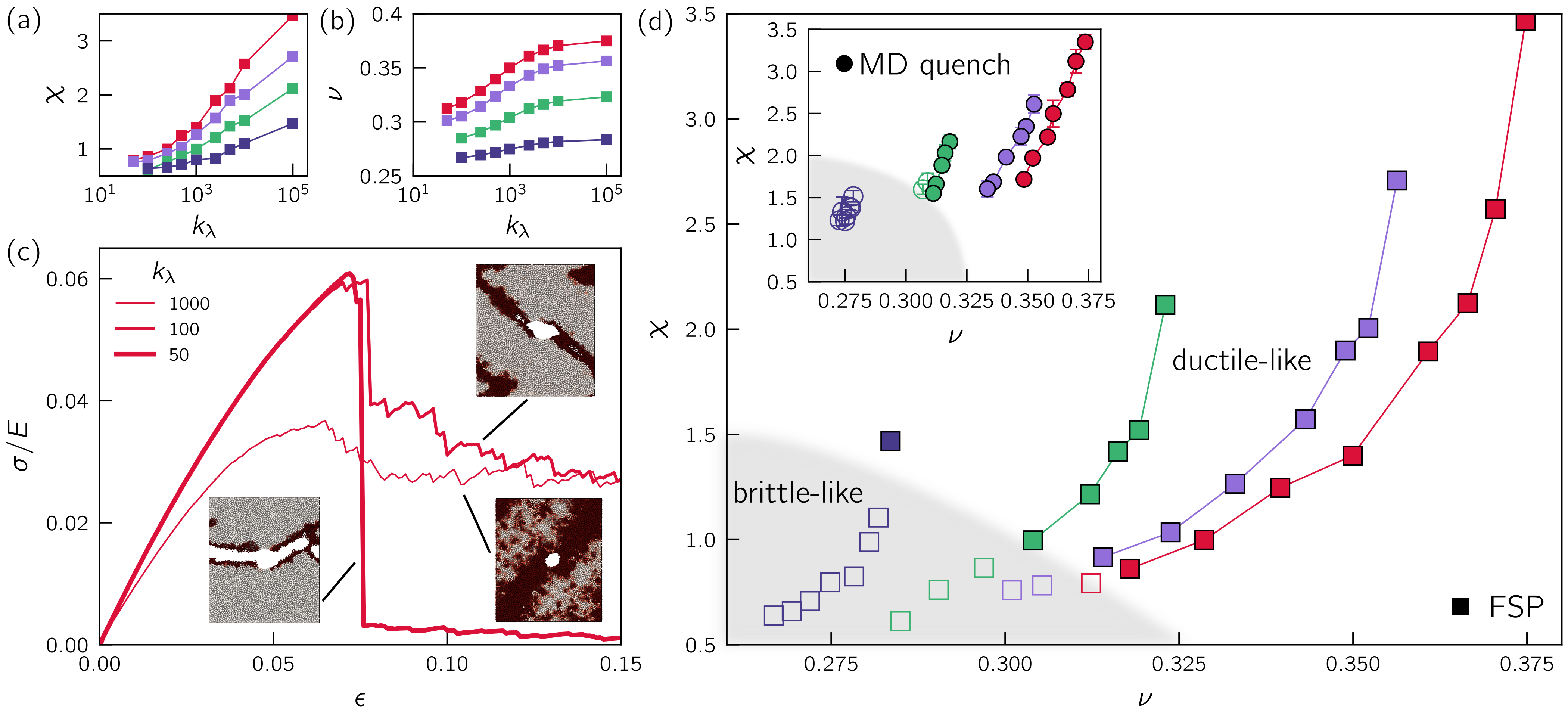}
  \caption{{\bf Brittle-ductile phase diagram in the $\chi$-$\nu$ plane under unconstrained uniaxial tension}. (a) $\chi$ vs.~the stiffness parameter $k_\lambda$ of FSP glasses. (b) The corresponding $\nu(k_\lambda)$. (c) Unconstrained uniaxial tension stress-strain curves for glasses with $r_{\rm c}=1.5$, prepared with different stiffness $k_\lambda$ as indicated in the legend. Snapshots show the non-affine deformation at $15\%$ strain. See text for additional discussion. (d) Ductile-brittle phase diagram in the $\chi\!-\!\nu$ plane for FSP glasses, where ductile-like glasses correspond to full symbols and brittle-like to empty ones. (inset) the same, but for glasses prepared via a conventional MD quench. The results presented in the inset correspond to those already presented in Figs.~\ref{fig:fig3}c-d.}
  \label{fig:fig4}
\end{figure*}

What additional physical quantities/factors/processes, which also vary with $\dot{T}$, play a role in ductile-to-brittle transitions in glasses? These, once identified, should allow to differentiate between the qualitatively different behaviors observed among the iso-$\!A_{\rm g}$ glasses, highlighted by open orange rectangles in Fig.~\ref{fig:fig3}a. As explained in the Introduction, there exists some evidence~\cite{schroers2004ductile,brittleness_BMG,castellero2007critical,greaves2011poisson,Greer_2012_AM,shi_2014,deng2018measuring}  that Poisson's ratio $\nu$
might affect the fracture toughness of glasses. The $\dot{T}$ dependence of $\nu$ for our glasses is presented in Fig.~\ref{fig:fig1}d, clearly indicating that iso-$\!A_{\rm g}$ glasses feature different $\nu$'s (marked by open orange squares). Consequently, we present in Fig.~\ref{fig:fig3}c a brittle-ductile phase diagram in the $A_{\rm g}\!-\!\nu$ plane, where ductile-like glasses correspond to full symbols and brittle-like to empty ones (we include data for all of our glasses, across the full range of $\dot{T}$ and $r_{\rm c}$ values employed, as in Fig.~\ref{fig:fig1}). It is observed that iso-$\!A_{\rm g}$ glasses of relatively high $\nu$ are ductile-like, while those with smaller $\nu$ are brittle-like. This indicates that $\nu$ might indeed plays a role in the fracture toughness of glasses.

What is the physical (and causal) effect of $\nu$ on the fracture toughness? Before setting out to address this important question, we aim at further substantiating the claim that both glassy disorder --- as quantified by either $A_{\rm g}$ or $\chi$ --- and some effect that is related to $\nu$ play a role in determining the fracture toughness of glasses. The results presented in Fig.~\ref{fig:fig3}c clearly support this claim, yet for relatively large values of $r_{\rm c}$ it is impossible to reach sufficiently small $A_{\rm g}$ levels to probe the brittle regime using conventional Molecular Dynamics (in which glasses are formed by quenching their corresponding liquids). Consequently, we employ below recently introduced computer algorithms that allow the generation of glassy solids featuring $A_{\rm g}\!\to\!0$, qualitatively representing extreme supercooling conditions.

To this aim, we first note that as shown in Fig.~\ref{fig:fig3}d, glasses form by quenching a liquid feature a clear relation between $A_{\rm g}$ and $\chi$ (see also \emph{Supplementary material}). The existence of such a relation, which appears to be weakly dependent on $r_{\rm c}$, suggests that for these glasses one can approximately use $A_{\rm g}$ and $\chi$ interchangeably. This is not the case for systems featuring $A_{\rm g}\!\to\!0$. The latter can be generated by a protocol that is referred to in what follows as the \emph{FSP algorithm}, described in detail in \emph{Methods} and in~\cite{fsp}; forming a computer glass using the FSP algorithm amounts to minimizing an augmented potential parameterized by a stiffness $k_\lambda$ that controls the mechanical noise of the resulting glasses (while $r_{\rm c}$ can still be varied, see \emph{Methods}), with lower-$k_\lambda$ FSP glasses featuring less mechanical fluctuations~\cite{fsp}. As is now established, FSP glasses can feature a gap in their nonphononic vDOS~\cite{fsp} for sufficiently small $k_\lambda$, and in particular in this limit they do not feature the $\omega^4$ law. That is, lower-$k_\lambda$ FSP glasses essentially feature $A_{\rm g}\!\to\!0$. Yet, these glassy solids still exhibit a finite and well-defined $\chi$, as demonstrated in Fig.~\ref{fig:fig4}a. Moreover, $\nu$ also varies systematically with $k_\lambda$ and $r_{\rm c}$, as shown in Fig.~\ref{fig:fig4}b.

The merit of FSP glasses in the present context is evident from panels (a) and (b) of Fig.~\ref{fig:fig4}. That is, we have at hand glasses with $A_{\rm g}\!\to\!0$ (in the low $k_\lambda$ regime), whose $\chi$ and $\nu$ can be systematically varied. Consequently, FSP glasses are most suitable for exploring the ductile-to-brittle transition in terms of both $\chi$ and $\nu$, providing access to  broader region of the brittle-like phase. In Fig.~\ref{fig:fig4}c, we present the stress-strain curves of FSP glasses with $r_{\rm c}\=1.5$ for three value of $k_\lambda$, together with spatial snapshots of each glass at a late stage in the loading process. For the largest $k_\lambda$ value ($k_\lambda\=1000$), the behavior is ductile-like, exhibiting extensive (system covering) plastic deformation and steady-state flow, essentially being insensitive to the initial crack. For the smallest $k_\lambda$ value ($k_\lambda\=50$), we enter into the brittle-like regime, where catastrophic failure takes place, accompanied by localized plastic deformation near the crack (which is perpendicular to the tensile axis) and an abrupt drop in the stress when the sample loses its load-bearing capacity. Interestingly, for the intermediate $k_\lambda$ value ($k_\lambda\=100$), the initial crack started growing along an inclined shear-band (resulting in a finite abrupt stress drop), but eventually plastic deformation in the shear-band took over (resulting in a gradual stress relaxation towards a plateau), with no catastrophic failure.
\begin{figure*}[t!]
  \includegraphics[scale=1]{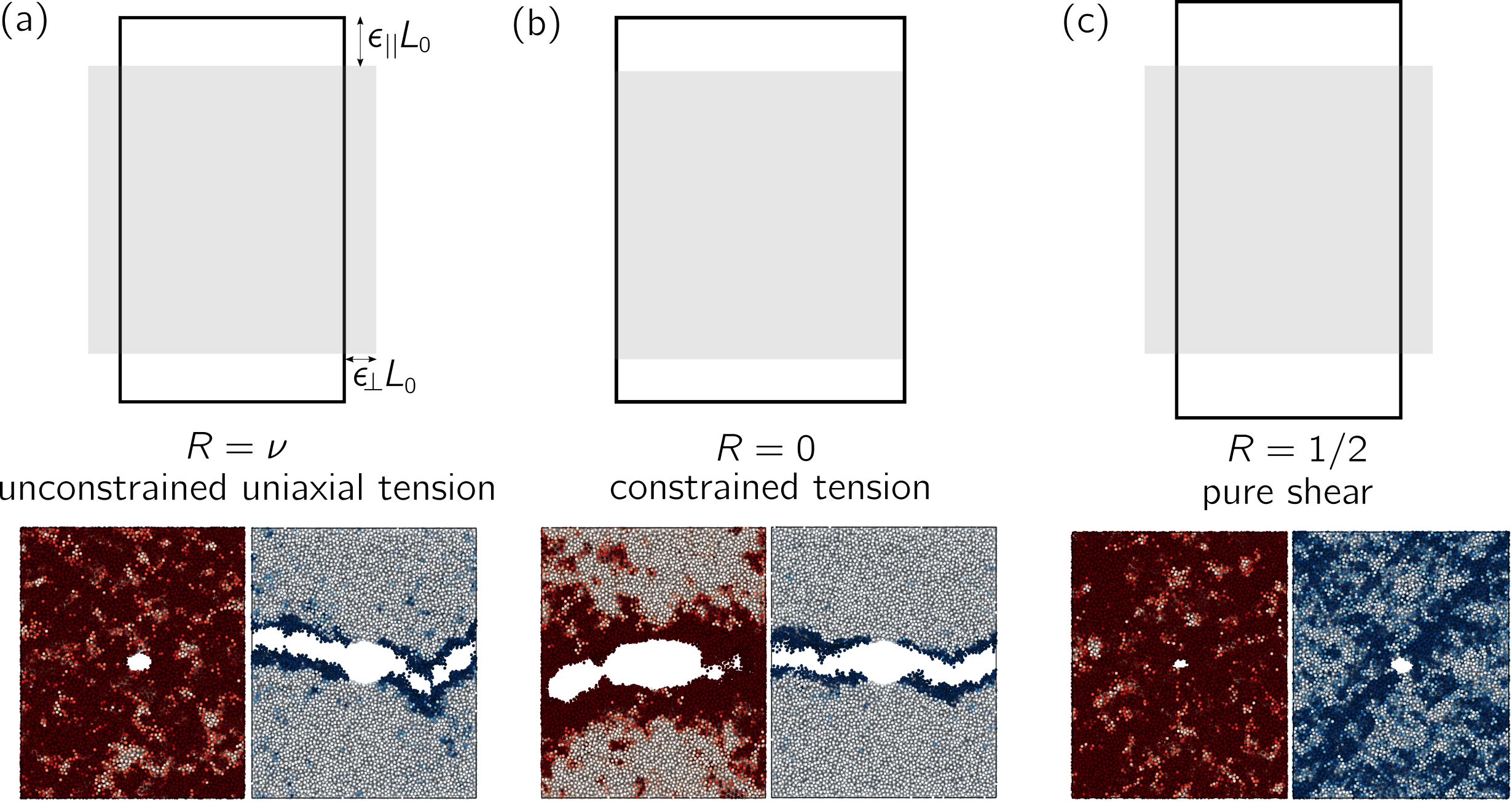}
  \caption{{\bf Different loading geometries and their effect on the imposed deformation}. (a) An illustration of a 2D projection of a 3D system (top), where the grey square of dimensions $L_0\!\times\!L_0$ represents the undeformed state. Under unconstrained uniaxial tension loading (applied strain $\epsilon_\|$ in the vertical direction), the system is free to contract in the two transverse/lateral directions (the out-of-plane direction is not shown), resulting in strain of magnitude $\epsilon_\perp\!<\!0$, see the empty rectangle (illustrating the deformed state). In this loading geometry, one has $R\!=\!-\epsilon_\perp/\epsilon_\|\!=\!\nu\!\le\!1/2$, i.e.~the relative importance of tensile and shear deformation is determined in this loading geometry by Poisson's ratio $\nu$. (b) The same as panel (a), but for constrained tension loading. In this loading geometry, one has $\epsilon_\perp\!=\!0$, resulting in $R\!=\!-\epsilon_\perp/\epsilon_\|\!=\!0$ (see empty rectangle, and note that the grey $L_0\!\times\!L_0$ square is the same as in panel (a)). (c) The same as panels (a) and (b), but for pure shear loading. In this loading geometry, one has $\epsilon_\perp\!=\!-1/2\epsilon_\|$, resulting in $R\!=\!-\epsilon_\perp/\epsilon_\|\!=\!1/2$ (see empty rectangle, and note that again the grey $L_0\!\times\!L_0$ square is the same as in panels (a) and (b)). At the bottom of each panel we present a pair of $\epsilon_\|\!=\!0.15$ snapshots, with $\nu\!=\!0.367$ (red, left) and $\nu\!=\!0.275$ (blue, right), deformed under the stated $R$ value. It is observed that the smaller $R$, the more localized plastic deformation is, and correspondingly the more brittle-like failure is.}
  \label{fig:fig5}
\end{figure*}

We performed an extensive analysis of all FSP glasses at hand (the four $r_{\rm c}$ values studied earlier and a broad range of $k_\lambda$ values) and present the results in Fig.~\ref{fig:fig4}d in a phase-diagram in the $\chi$-$\nu$ plane, where a ductile-like behavior corresponds to full squares (color code as in Fig.~\ref{fig:fig1}) and a brittle-like behavior corresponds to empty squares. First, it is observed that indeed FSP glasses allow to explore a broader region of the brittle-like phase. Second, and most importantly, it is observed that both $\chi$ and $\nu$ (more precisely some effect that is related to $\nu$) play a role in the ductile-to-brittle transition, as is evident from the curved phase boundary (a blurry region is added as a guide to the eye).

Finally, we add in the inset the results for the glasses formed by quenching a liquid --- presented earlier in Fig.~\ref{fig:fig3} (but not in the $\chi$-$\nu$ plane). Interestingly, despite the widely different glass formation protocols involved, the two classes of glasses appear to reveal a qualitatively similar ductile-brittle phase diagram in the $\chi$-$\nu$ plane, where both decreasing $\chi$ and $\nu$ promote a brittle-like behavior. It is important to note again that the results presented in Fig.~\ref{fig:fig4}d were obtained for our choice of initial crack of fixed geometry and dimensions (see also \emph{Supplementary material}); varying the latter may change quantitative aspects of the results, but the combined effect of $\chi$ and $\nu$ on the fracture toughness under unconstrained uniaxial tension conditions is general. Once this is established, we focus our attention on the question posed above regarding the physical (and causal) effect of $\nu$ on the fracture toughness, which is addressed next.

\subsection{The interplay of loading geometry and plastic deformability: Constrained tension and soft defects again}

Poisson's ratio $\nu$ is a linear elastic (reversible) response coefficient that quantifies the relative transverse contraction of a material under unconstrained (transversely-free) uniaxial tension. In assessing the relevance of $\nu$ to the fracture toughness of glasses, one can a priori distinguish two qualitatively different classes of physical effects. First, $\nu$ can be viewed as a directly and easily measurable quantity that allows to indirectly probe nonlinear irreversible response coefficients that are relevant for plastic deformability and toughness. Indeed, $\nu$ is expected to vary when the material structure and interactions vary, and in particular it exhibits systematic variations with $A_{\rm g}$ and $\chi$, as demonstrated in Figs.~\ref{fig:fig3}-\ref{fig:fig4}. Second, $\nu$ --- as a linear elastic coefficient --- may affect the geometry of deformation glasses experience under external driving forces. In particular, in the unconstrained uniaxial tension tests exclusively considered up to now, $\nu$ controls the degree of Poisson contraction, which in turn determines the relative magnitude of tensile and shear deformation. That is, $\nu$ in this loading geometry will affect the activation of soft defects (whose number is controlled by $A_{\rm g}$), which are generally more sensitive to shear deformation.

As physically relevant disorder measures such as $A_{\rm g}$ and $\chi$ have been identified and quantified, we hypothesize that a linear elastic response coefficient such as $\nu$ does not contain \emph{additional} physical information about plastic deformability, on top of $A_{\rm g}$ and/or $\chi$. Consequently, while in the absence of well-founded quantifiers of glassy disorder one may use $\nu$ as some rough proxy of disorder, when these are at hand and taken into account --- as is the case here --- the effect of $\nu$ in the phase diagram of Fig.~\ref{fig:fig4}d strongly points towards the second possibility discussed above. That is, one needs to understand the role of the deformation/loading geometry on the fracture toughness, both in relation to $\nu$ and in more general terms.

To address the latter, we introduce here the biaxiality ratio $R$ (cf.~Fig.~\ref{fig:fig5}), defined as $R\!\equiv\!-\epsilon_\perp/\epsilon_\|$, where $\epsilon_\|$ is the tensile/extensional strain and $\epsilon_\perp$ corresponds to the transverse strains, i.e.~the strain components in the directions perpendicular to the applied tension axis. The main utility of $R$ is that it quantifies the relative importance of tensile and shear deformation imposed by the loading geometry, and since soft defects (STZs) are most sensitive to shear deformation~\cite{falk_langer_stz}, $R$ offers a simple measure of the fraction of soft defects being activated under a certain loading geometry (while their number is controlled by $A_{\rm g}$). Since, in principle, $\epsilon_\|$ and $\epsilon_\perp$ can be independently controlled, $R$ can take any value (and can even evolve during a mechanical test). Yet, we focus here on a few cases that are most relevant to our discussion. For pure shear deformation, defined by the absence of linear dilatational strain ($\epsilon_\|+2\epsilon_\perp\=0$), we have $R\=1/2$. As plastic deformation in glasses is most sensitive to shearing~\cite{falk_langer_stz}, this value of $R$ is expected to facilitate a more plastic response, see Fig.~\ref{fig:fig5}c. For the unconstrained uniaxial tension loading geometry considered up to now, we have $R\=\nu\!\le\!1/2$; that is, $R$ in this loading geometry is determined by a material property, quantified by $\nu$. Therefore, we see that reducing $\nu$ in the unconstrained loading configuration tends to increase the relative importance of tensile deformation (cf.~Fig.~\ref{fig:fig5}a), which may facilitate more brittle-like behavior, in qualitative agreement with the results of Fig.~\ref{fig:fig4}d.

As $R$ is in general not a material property, but can rather be controlled through the imposed loading geometry, one can use it to disentangle the roles of $A_{\rm g}$ (or $\chi$) --- i.e.~the role of soft defects --- and of the loading geometry on the fracture toughness. To that aim, we performed constrained tension tests, where the transverse/lateral boundaries (perpendicular to the main tensile axis) are held fixed, i.e.~$\epsilon_\perp\=0$. This case corresponds to $R\=0$, see Fig.~\ref{fig:fig5}b and \emph{Methods} for additional discussion. The loading geometry corresponding to $R\=0$ allows to eliminate the role of $\nu$ as a linear response coefficient that controls Poisson contraction. Consequently, we expect the fracture toughness in this case to be determined by $A_{\rm g}$. Since for $R\=0$, which obviously favors tensile deformation, catastrophic failure is unavoidable, $A_{\rm g}$ is expected to affect the value of the fracture toughness, not strictly a brittle-to-ductile transition. In particular, we expect iso-$\!A_{\rm g}$ glasses to feature very similar toughness under constrained tension, despite having different $\nu$'s, and the toughness to systematically increase with $A_{\rm g}$ for the same glass cooled at different rates $\dot{T}$.

\begin{figure}[t!]
  \includegraphics[scale=1]{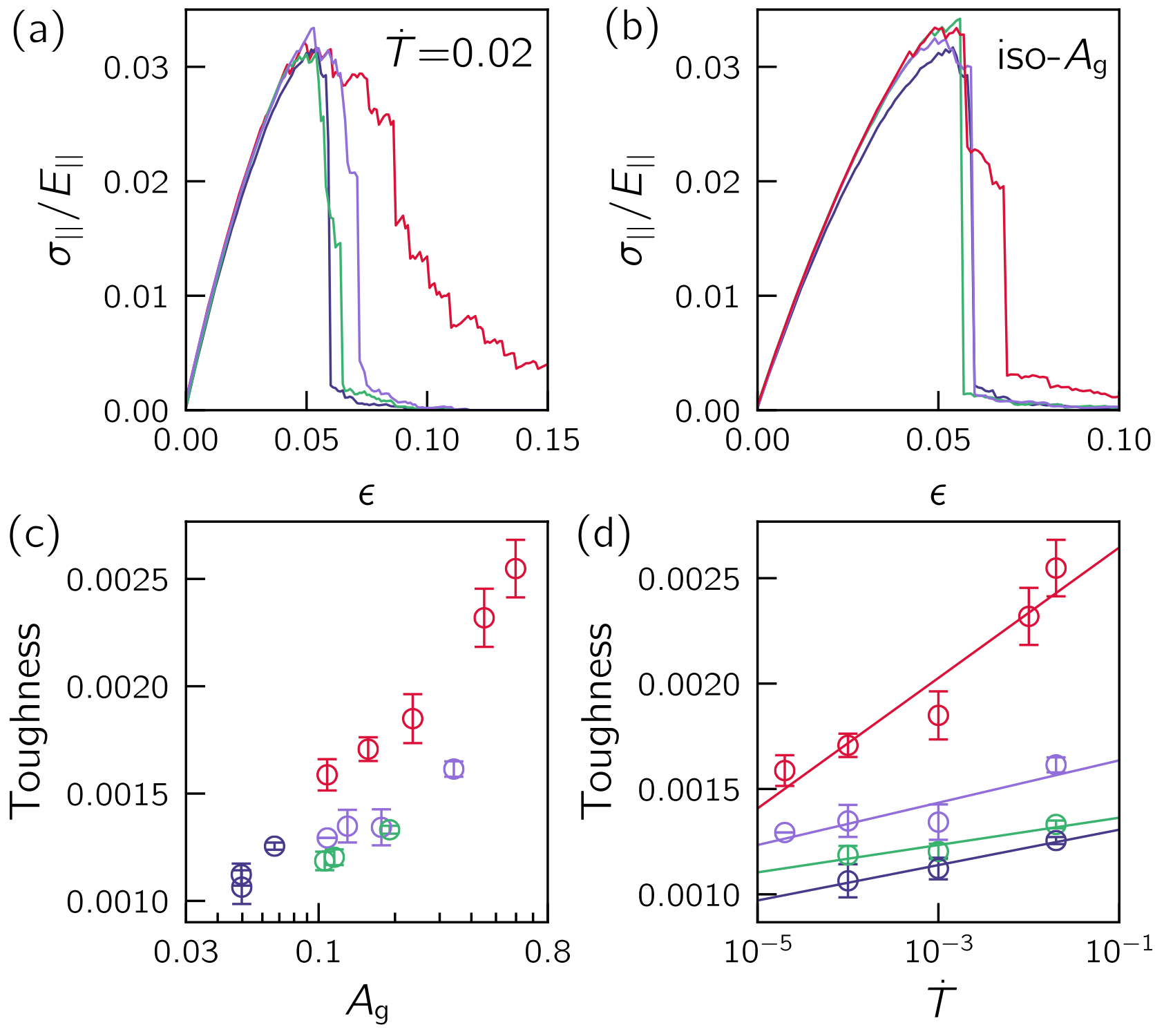}
  \caption{{\bf Disentangling the effect of soft defects and loading geometry on the toughness}. (a) Stress-strain curves under constrained tension, for glasses prepared at a high quench rate $\dot{T}\!=\!0.02$ and different $r_{\rm c}$'s. See text and \emph{Methods} for the exact definitions of $\sigma_\|$, $\epsilon_\|$ and $E_\|$. (b) The same as in panel (a), but for iso-$\!A_{\rm g}$ glasses, featuring approximate collapse. (c) The Toughness (computed as the integral over the stress-strain curve and normalized by $E_\|$, see text for details) vs.~$A_{\rm g}$. (d) The Toughness vs.~$\dot{T}$.}
  \label{fig:fig6}
\end{figure}

These predictions are tested in Fig.~\ref{fig:fig6}, where we first show in panel (a) that different glasses (i.e.~different $r_{\rm c}$ values) feature rather large variability in their failure behavior under constrained tension for a fixed cooling rate $\dot{T}$ (here we present stress-strain curves; a quantitative measure of the fracture resistance is presented in panels (c) and (d)). The discussion above predicts that if one repeats the calculations of Fig.~\ref{fig:fig6}a for different glasses featuring nearly the same $A_{\rm g}$, under the same constrained tension loading geometry ($R\=0$), most of the variability in the failure behavior would be gone. The results of these calculations are presented in Fig.~\ref{fig:fig6}b, where it is indeed observed that the stress-strain curves of iso-$\!A_{\rm g}$ glasses featuring different $\nu$ values, approximately collapse, as predicted. These results should be compared to --- and contrasted with --- the results shown in Fig.~\ref{fig:fig3}c obtained under unconstrained uniaxial tension. We stress that these results strongly suggest that once the geometrical Poisson contraction is eliminated (or more generally when $R$ is prescribed), Poisson's ratio $\nu$ does not play a major role in determining the toughness of glasses.

Another prediction discussed above is that the fracture toughness is expected to systematically increase with $A_{\rm g}$ under constrained tension. In order to test this prediction, we quantify the fracture resistance through the ``Toughness''~\cite{meyers2008mechanical}, which is defined as the integral under the stress-strain curve up to failure (it has the dimensions of energy density, and should be distinguished from the ``Fracture toughness'' $K_{\rm c}$~\cite{lawn1993fracture}, which has the dimensions of stress times the square root of length, see \emph{Supplementary material}). Note that the Toughness is usually employed for samples that do not contain an initial crack (as it obviously depends on the crack geometry and dimensions); here we do apply it to such samples, but as done throughout this work, the initial crack geometry and dimensions are strictly kept fixed. In Fig.~\ref{fig:fig6}c, we plot the Toughness vs.~$A_{\rm g}$ for our entire glasses ensemble, calculated under constrained tension. As predicted, we observe systematic increase of the Toughness with $A_{\rm g}$. Finally, we expect that the $\dot{T}$ dependence of $A_{\rm g}$, previously presented in Fig.~\ref{fig:fig1}f, is fully and transparently mirrored in the dependence of the Toughness on $\dot{T}$. This is strongly supported by the results shown in Fig.~\ref{fig:fig6}d.

\section{Discussion and conclusion}

In this work, employing extensive atomistic simulations and recently developed concepts, we studied the physical origin of the failure resistance of glasses, and in particular the emerging brittle-to-ductile transitions. We showed that these are controlled by both the abundance of soft defects --- as quantified by $A_{\rm g}$, the prefactor of the universal $\omega^4$ vDOS of nonphononic excitations in glasses (and indirectly by $\chi$) --- and by the loading geometry of the fracture test employed to extract the toughness. The loading geometry/configuration is shown to affect the relative magnitude of shear and tensile deformation experienced by the material near the initial crack, and consequently the emerging plastic dissipation, for a fixed $A_{\rm g}$. Roughly speaking, $A_{\rm g}$ controls the number of soft defects/STZs in a glass, and the loading geometry controls the fraction of soft defects/STZs being actually activated, and hence both affect plastic deformation and consequently the fracture toughness.

We find that only under a certain choice of loading geometry, where Poisson contraction can take place, Poisson's ratio $\nu$ can be sensibly used to quantify the fracture toughness \emph{together} with $A_{\rm g}$ (or $\chi$). These results suggest that brittle-to-ductile transitions in glasses are not controlled by a critical Poisson's ratio, as previously proposed, and elucidate the physical role $\nu$ might play in affecting the fracture toughness of glasses.

Our findings thus provide basic insights into the physical origin of the failure resistance of glasses, and should serve as important input for theories that aim at predicting it. Such theories should include the number of soft defects, as quantified here by $A_{\rm g}$, and its deformation-induced spatiotemporal evolution as important ingredients. Our findings also call for additional future investigations, along the lines we delineate next.

In this work we employed an athermal ($T\!=\!0$) and quasistatic (i.e.~we approach the limit of vanishing deformation rates) glass-deformation protocol~\cite{Malandro_Lacks,lemaitre2004}. While this choice is suitable for identifying structural and loading geometry effects on the fracture toughness, it completely suppresses thermal and rate effects, and might therefore introduce uncontrolled  artefacts. It is obviously desirable to extend the present study to more physically realistic loading protocols that involve both finite temperatures and strain-rates.

Equally important for future work are mechanical tests on computer glasses that trace out the role of sample size and thickness in determining the failure resistance to failure; in this work we employed glassy slabs of small thickness (see \emph{Methods}), hence systematically studying possible thickness effects is in place. In the \emph{Supplementary material} we show that increasing the glass in-plane dimensions --- other than the slab thickness --- does not appear to affect the fracture behavior. In addition, as our results were obtained for an initial crack of fixed geometry and dimensions (but see also \emph{Supplementary material}), it would be important in future work to sufficiently increase both the system size and the initial crack size to obtain quantitative results that are entirely independent of both.

We selected to study a rather flexible computer glass model system --- introduced first in~\cite{itamar_brittle_to_ductile_pre_2011} --- that features a very large variability of glasses' mechanical properties~\cite{sticky_spheres_part_1,sticky_spheres_part_2}. While we are unable at this point to single out a particular deficiency of this model, the employed pairwise interactions might be too simple to capture the emergent effects of more complex and realistic particle interactions of laboratory glasses in their entirety. For example, the Poisson ratio of our computer glasses does not fall below $\nu\!=\!1/4$, while other (network) computer glasses (e.g.~those employed in~\cite{shi_2014}) can feature $\nu\!<\!1/4$.

At the same time, the power of computer simulations allows us to probe and measure generally applicable quantifiers of mechanical disorder such as $A_{\rm g}$ and $\chi$; while the former does not appear to be directly accessible experimentally, the latter could be --- in principle --- indirectly accessed in future experiments via transverse sound attenuation measurements at wavenumbers $\lesssim\!1$nm$^{-1}$, or via other experimental techniques~\cite{huo2013dependence,zhu2017correlation}. As new and more informative quantifiers of STZ-densities are being developed (cf.,~e.g.,~\cite{david_huge_collaboration,zohar_prerc,pseudo_harmonic_modes_prl_2021}), we cannot rule out that more informative dimensionless quantifiers of mechanical disorder can shed further light on the relation between mechanical disorder and plastic deformability, and hence on the fracture toughness.

In our analysis of the abundance of STZs as captured by $A_{\rm g}$, we have not considered the effect of geometric/mechanical coupling of STZs to different deformation modes. In~\cite{sticky_spheres_part_1} it was shown that varying $r_{\rm c}$ can dramatically affect the ratio of dilatational to shear eigenstrains associated with STZs viewed as Eshelby inclusions. These changes in the intrinsic geometric properties of soft defects can potentially play a role in the failure of glasses. For example, the coupling of soft defects to different deformation modes can affect the susceptibility to cavitation, which is in fact observed in our simulations (but will be discussed elsewhere) as predicted in~\cite{Rycroft2012}.

Finally, our work constitutes a step towards a quantitative understanding of deformation-induced failure in disordered solids. Future investigations should aim at better understanding the role of deformation-induced rejuvenation of glassy structures --- going beyond $A_{\rm g}$ that characterizes the \emph{initial} glass structure --- in controlling the ductile or brittle nature of failure. That is, plastic deformation itself could be limited --- in some cases --- in its ability to generate new soft defects that facilitate further plastic flow in the material, as typically happens within shear bands~\cite{barbot2020rejuvenation}.

\section{Materials and Methods}

Here we provide details about the computer glass models employed, about the methods used to deform the glass samples, and about the observables considered.

\subsection{Computer glass models}

We employ a glass model put forward in~\cite{itamar_brittle_to_ductile_pre_2011}; in this model, particles of equal mass $m$ interact via the following pairwise potential
\begin{widetext}
\begin{equation}
    \varphi(r_{ij},\lambda_{ij}) \!=\!
\left\{
\begin{array}{cc}
\!\!4\varepsilon \bigg[ \big(\frac{\lambda_{ij}}{r_{ij}}\big)^{12} - \big(\frac{\lambda_{ij}}{r_{ij}}\big) ^{6} \bigg],
     &  \frac{r_{ij}}{\lambda_{ij}} <  x_{\mbox{\tiny min}}  \\
\varepsilon \bigg[a\big(\frac{\lambda_{ij}}{r_{ij}}\big)^{12} -b\big(\frac{\lambda_{ij}}{r_{ij}}\big)^{6} + \sum\limits_{\ell=0} ^{3}  c_{\mbox{\tiny $2\ell$}} \big(\frac{r_{ij}}{\lambda_{ij}}\big)^{2\ell} \bigg] , & x_{\mbox{\tiny min}}\!\le\! \frac{r_{ij}}{\lambda_{ij}}< x_c\\
0\,,  & x_c \le \frac{r_{ij}}{\lambda_{ij}}
\end{array}
\right. ,
 \label{eq:potential}
\end{equation}
\end{widetext}
where $\varepsilon$ is a microscopic energy scale, $x_{\mbox{\tiny min}},x_c$ are the (dimensionless) locations of the minimum of the Lennard-Jones potential and modified cutoff, respectively, and the $\lambda_{ij}$'s are the length parameters, described further below. We express the dimensionless cutoff $x_c$ in terms of $x_{\mbox{\tiny min}}\!=\!2^{1/6}$, for simplicity, by defining $r_c\!\equiv\!x_c/x_{\mbox{\tiny min}}$. $r_c$ serves as one of the key control parameters in our study; we refer readers to~\cite{sticky_spheres_part_1,sticky_spheres_part_2} for a comprehensive study of the \emph{elastic} properties of glasses formed with this model, under variations of $r_c$ and of glass preparation protocol. The coefficients $a,b,\{c_{\mbox{\tiny $2\ell$}}\}$ are chosen such that the attractive and repulsive parts of $\varphi$, and its first two derivatives, are continuous at $x_{\mbox{\tiny min}}$ and at $x_c$, see  Table~\ref{tab:potential_coefficients} for the coefficients' numerical values. Unless specified otherwise, we employ simulational units, where energies are expressed in terms of $\varepsilon$, temperature in terms of $\varepsilon/k_B$, lengths in terms of $\lambdabar$ (see further discussion below), elastic moduli in terms of $\epsilon/\lambdabar^3$, and times in terms of $(m\lambdabar^2/\varepsilon)^{1/2}$.

\begin{table*}
\caption{\label{tab:potential_coefficients}
Pairwise potential coefficients}
\begin{ruledtabular}
\begin{tabular}{ccccc}
coefficient & $r_{c} = 1.2$ & $r_{c} = 1.3$ & $r_{c} = 1.4$ & $r_{c} = 1.5$ \\
\hline
a & -106.991613526652 & -17.7556513878655 & -2.942014535960528 & 1.1582440286928275 \\
b & -304.918469059567 & -50.37332289908061 & -12.11892507229410 & -2.2619482444770567  \\
$c_{\mbox{\tiny $0$}}$& -939.388037994211 & -138.58271673010657 & -35.72455291073821740 & -12.414700446492716 \\
$c_{\mbox{\tiny $2$}}$& 1190.70962256002 & 161.71576064627635 & 38.70071979329345996  & 12.584354590303674 \\
$c_{\mbox{\tiny $4$}}$& -541.3001315875512 & -66.7252832098764 & -14.5415594738601088 & -4.320508006050397 \\
$c_{\mbox{\tiny $6$}}$& 85.86849369147127 & 9.50283097488097 & 1.86201465049568722 & 0.49862551162881885 \\
\end{tabular}
\end{ruledtabular}
\end{table*}

We employ the pairwise potential in Eq.~(\ref{eq:potential}) in two distinct protocols for creating glasses. For the first procedure, we follow the conventional computer glass formation route: high-temperature equilibrium liquid configurations at pressure $P\!=\!1$ are generated at an initial temperature $T_{\mbox{\tiny initial}}\!=\!0.7$. Those equilibrium configurations are then quenched at fixed zero pressure $P\!=\!0$, and at a constant cooling rate $\dot{T}$ (as specified in the figure legends) until a temperature $T_{\mbox{\tiny final}}\!=\!0.05$ is reached. Any residual heat is subsequently removed using a standard energy minimization algorithm. For this protocol, we employ a 50:50 binary mixture of `large' (l) and `small' (s) particles, and fix the effective size parameters $\lambda_{\rm ss}\!=\!\lambdabar$, $\lambda_{\rm sl}\!=\!\lambda_{\rm ls}\!=\!1.18\lambdabar$, and $\lambda_{\rm sl}\!=\!1.4\lambdabar$, for small-small, small-large, and large-large interactions, respectively.

Glasses formed with the second procedure --- introduced next --- are referred to as \emph{Fluctuating-Size-Particles} (FSP) glasses. They were created by employing the algorithm put forward in~\cite{fsp}; within this algorithm, an augmented potential energy is employed for creating glasses with very high variability of their mechanical stability. This variability is achieved during glass formation by allowing particle sizes $\lambda_i$ to fluctuate about a preferred effective size $\lambda_i^{(0)}$ at an energetic penalty determined by a potential of the form
\begin{equation}
    \tilde{\varphi}(\lambda_i)=\frac{k_\lambda}{2}\big(\lambda_i-\lambda_i^{(0)}\big)^2\bigg[\bigg(\frac{\lambda_i^{(0)}}{\lambda_i}\bigg)^2+\frac{1}{4}\bigg] \ ,
\end{equation}
where the stiffness $k_\lambda$ constitutes the main control parameter of FSP glasses. Notice that this form is slightly different from the effective-size potential used in~\cite{fsp}; the latter turns out to be unstable for small $k_\lambda$ and small confining pressures. We set the preferred sizes of half the particles at $\lambda_i^{(0)}\!=\!\lambdabar/2$ (as before, $\lambdabar$ forms the microscopic units of length), the other half at $\lambda_i^{(0)}\!=\!7\lambdabar/10$, and used the convention $\lambda_{ij}\!=\!\lambda_i\!+\lambda_j$. Crucially, once a glass is created with the FSP algorithm, all subsequent analyses and mechanical tests are performed under \emph{fixed} particle sizes $\lambda_i$, namely using the exact same pairwise potential as given by Eq.~(\ref{eq:potential}). Importantly, local minima (glassy states in mechanical equilibrium) of the augmented potential also constitute local minima of the potential given by Eq.~(\ref{eq:potential}). Finally, we note that the particles of FSP glasses are polydispersed in size, since during glass formation the effective particle sizes fluctuate, see~\cite{fsp} for further discussion and details. We employ the FSP algorithm to create zero pressure glasses, starting from high temperature liquid states and minimizing the augmented potential using a variety of stiffnesses $k_\lambda$ (and various values of $r_c$) as reported in the figure legends. In general, decreasing the stiffness parameter $k_\lambda$ results in stiffer glasses featuring less mechanical fluctuations.

Using the two protocols described above we created two sets of ensembles: one set was created for extracting the micromechanical properties of each ensemble as reported in Fig.~\ref{fig:fig1} for glasses formed with a finite cooling-rate, and in Fig.~\ref{fig:fig4} for glasses formed using the FSP protocol. For these calculation we prepared between 256 to 5000 glasses of $N\!=\! 4096$ particles (except for lowest cooling rate $\dot{T}\!=\!2.10^{-5}$ for which $N\!=\! 2197$). The second set of ensembles --- prepared for our mechanical tests --- were larger, rectangular glass slabs of dimensions $L_x\!\times\!L_y\!\times\!L_z$, with $L_x\=L_y\=L_0=60a_0$ and $L_z\=15a_0$ (recall that $a_0\!\equiv\!(V/N)^{1/3}$), containing $N\!=\!54000$ particles in total.

\subsection{Mechanical loading simulations}

Mechanical loading is carried out via athermal quasistatic simulations, i.e. zero strain-rate and temperature. For both unconstrained uniaxial tension and constrained tension (see also Fig.~\ref{fig:fig5} and ``Mechanical and structural observables'' below), the glass is affinely deformed by imposing an extensional strain along the $y$-axis (regarded as the parallel direction in the text). We define the extensional strain as $\epsilon_\|\=(L-L_0)/L_0$, where $L$ and $L_0$ are the current and initial box length along $\vec{e}_y$, respectively. Particle positions are changed according to $y_i\!\to\! y_i+\delta\epsilon_\|\,y_i$, where the strain step $\delta\epsilon_\|$ is fixed at each step such that the accumulated extensional strain $\epsilon_\|$ increases by $10^{-3}$. Subsequently, we minimize the potential energy $U$, where periodic boundary conditions are applied in all directions. For constrained tension, we employ the conventional conjugate gradient algorithm. For unconstrained uniaxial tension, we combine the FIRE minimization algorithm~\cite{bitzek2006structural} with the Berendesen barostat. The latter keeps the stress in the transverse directions, i.e.~along the $x$- and $z$-axis, zero, with a time constant $\tau_{\rm Ber}\=10.0$. For both loading geometries, the minimization is stopped once the ratio between the typical gradient of the potential and the typical interparticle force drops below $10^{-10}$. During each simulations, and closely following Ref.~\cite{falk_langer_stz}, we monitor the non-affine part of the plastic deformation based on the commonly used $D^2_{\rm min}$ field~\cite{falk_langer_stz} between the initial configuration ($\epsilon_{\|}\=0$) and the current state.

\subsection{Mechanical and structural observables}

The potential energy of our glasses is given by $U\!=\!\sum_{i<j}\varphi_{ij}(r_{ij})$, where the pairwise potential we employed is described in Eq.~(\ref{eq:potential}). As we focus on athermal conditions, the shear modulus is defined as~\cite{lutsko}
\begin{equation}
    \mu \equiv \frac{1}{V}\left(\frac{\partial^2U}{\partial\gamma^2} - \frac{\partial^2U}{\partial\gamma\partial\xv}\cdot\calBold{H}^{-1}\cdot\frac{\partial^2U}{\partial\xv\partial\gamma} \right)\,,
\end{equation}
where
\begin{equation}\label{eq_hessian}
    \calBold{H}\equiv\frac{\partial^2U}{\partial\xv\partial\xv}
\end{equation} is the Hessian matrix of the potential $U$, and $\xv$ denotes particle coordinates. The latter are considered to transform via $\xv\!\to\!\bm{H}(\gamma)\cdot\xv$ with the parameterized shear transformation
\begin{equation}\label{shear_transformation_matrix}
\bm{H}(\gamma) =  \left( \begin{array}{ccc}1&\gamma&0\\0&1&0\\
0&0&1\end{array}\right)\,.
\end{equation}
The pressure is given by
\begin{equation}
p\equiv-\frac{1}{V\dbar}\frac{\partial U}{\partial \eta}\,,
\end{equation}
where $\dbar\!=\!3$ is the dimension of space, and $\eta$ is the isotropic dilatational strain that parameterizes the suitable transformation of coordinates $\xv\!\to\!\bm{H}(\eta)\cdot\xv$ as
\begin{equation}\label{dilation_transformation_matrix}
\bm{H}(\eta) =  \left( \begin{array}{ccc}e^\eta&0&0\\0&e^\eta&0\\0&0&e^\eta\end{array}\right)\,.
\end{equation}
The athermal bulk modulus $K\!\equiv -\frac{1}{\dbar}\frac{dp}{d\eta}$ is given by
\begin{equation}\label{eq-K}
    K= \frac{1}{V\dbar^2}\left(\frac{\partial^{2}U}{\partial \eta^{2}} -\dbar \frac{\partial U}{\partial\eta}- \frac{\partial^{2}U}{\partial \eta \partial \xv} \cdot \calBold{H}^{-1} \cdot \frac{\partial^{2}U}{\partial\xv\partial\eta}\right)\,.
\end{equation}
With the definitions of the shear and bulk moduli in hand, the Poisson's ratio $\nu$ of a 3D solid is given by
\begin{equation}\label{eq-poisson}
    \nu \equiv \frac{3K-2\mu}{6K+2\mu}\,.
\end{equation}

As explained above (see ``Mechanical loading simulations'') and in the main text, we employed two different types of loading geometries, namely unconstrained uniaxial tension and constrained tension (cf.~Fig.~\ref{fig:fig5}). In the former, uniaxial tension of magnitude $\sigma_\|$ is applied, while keeping the transverse stresses (in the two perpendicular directions) zero, i.e.~$\sigma_{\perp}\!=\!0$. The resulting extensional strain $\epsilon_\|$ satisfies $\sigma_\|\=E\epsilon_\|$, where $E\=2\mu(1+\nu)$
is the conventional Young's modulus. In this loading geometry, the biaxiality ratio is set by Poisson's ratio, i.e.~$R\=-\epsilon_\perp/\epsilon_\|\=\nu$.

In the latter case, i.e.~in the constrained tension loading configuration (cf.~Fig.~\ref{fig:fig5}), the transverse boundaries (perpendicular to the applied tension direction) are held fixed, i.e.~$\epsilon_\perp\=0$, and consequently the transverse stresses $\sigma_{\perp}$ are no longer zero (in fact, they are tensile, i.e.~$\sigma_{\perp}\!>\!0$). Note that since $\sigma_{\perp}\!\ne\!0$, i.e.~the resulting state of stress is no longer uniaxial (though uniaxial stress $\sigma_\|\!>\!0$ is applied), we term this loading geometry `constrained tension', omitting `uniaxial'). In this loading geometry, one can define two different moduli (linear response coefficients) $E_\|$ and $E_{\perp}$, according to $\sigma_\|\=E_\|\epsilon_\|$ and $\sigma_\perp\=E_\perp\epsilon_\perp$. Using Hooke's law, one finds
\begin{equation}
E_\| = 2\mu\,\frac{1-\nu}{1-2\nu}\qquad\hbox{and}\qquad E_{\perp} = 2\mu\,\frac{\nu}{1-2\nu} \ .
\end{equation}
These expressions are verified against numerical simulations in the \emph{Supplementary material}. Note that $E_\|(\mu,\nu)\!\ge\!E(\mu,\nu)$ for every $\mu$ and $\nu$, and that in this case we have a zero biaxiality ratio, i.e.~$R\=-\epsilon_\perp/\epsilon_\|\=0$.

\subsection{Quantifiers of mechanical disorder}

In our work we consider two dimensionless quantifiers of mechanical disorder. The first quantifier is defined as
\begin{equation}
    \chi \equiv \frac{\Delta\mu}{\langle\mu\rangle}\sqrt{N}\,,
\end{equation}
where $\Delta\mu$ is the ensemble-standard-deviation of the shear modulus $\mu$ (of glasses created with the exact same protocol), $N$ is the system size, and $\langle\mu\rangle$ denotes the ensemble-average shear modulus. $\chi$ has been studied in~\cite{sticky_spheres_part_1} under variations of glasses' interparticle interaction potential, and under different glass annealing protocols in~\cite{scattering_jcp_2020,minimally_disordered_glasses_arXiv}.

The second quantifier of mechanical disorder considered in our work is the prefactor $A_{\rm g}$ of the nonphononic spectrum of a glass, which is of the form ${\cal D}(\omega)\!=\!A_g\omega^4$, independent of spatial dimension~\cite{modes_prl_2018} or microscopic details~\cite{modes_prl_2020}. The latter is obtained by performing a partial diagonalization of the Hessian matrix $\calBold{H}$, defined in Eq.~(\ref{eq_hessian}), and calculated for each member of our glass ensembles, to obtain the eigenfrequencies $\omega_\ell$ that solve the eigenvalue problem $\calBold{H}\cdot\bm{\psi}_\ell\!=\!\omega^2_\ell\,\bm{\psi}_\ell$ (all masses are all set to unity). We then histogram the eigenfrequencies for each glass ensemble, to obtain their distribution ${\cal D}(\omega)$ over the frequency axis. Importantly, we note that the scaling form of ${\cal D}(\omega)\!\sim\!\omega^4$ is most readily observable in glasses of sizes small enough to suppress the otherwise-dominant phononic waves that emerge in the low-frequency spectrum of any elastic solid, as explained in detail in~\cite{modes_prl_2016}.\\

{\bf Acknowledgments and declaration of conflicts and other disclosures} D.R.~acknowledges support of the Simons Foundation for the
``Cracking the Glass Problem Collaboration'' Award No.~348126. E.L.~acknowledges support from the NWO (Vidi grant no.~680-47-554/3259). E.B.~acknowledges support from the Ben May Center for Chemical Theory and Computation and the Harold Perlman Family.\\


%

\onecolumngrid
\newpage
\begin{center}
	\textbf{\large Supplementary material: ``Brittle to ductile transitions in glasses: Roles of soft defects and loading geometry''}
\end{center}

The goal of this document is to provide additional supporting data to the results presented in the manuscript, as detailed in the table of contents.

\setcounter{equation}{0}
\setcounter{figure}{0}
\setcounter{section}{0}
\setcounter{table}{0}
\setcounter{page}{1}
\makeatletter
\renewcommand{\theequation}{S\arabic{equation}}
\renewcommand{\thefigure}{S\arabic{figure}}
\renewcommand{\thesubsection}{S-\arabic{subsection}}
\renewcommand{\thesection}{S-\arabic{section}}
\renewcommand*{\thepage}{S\arabic{page}}
\renewcommand{\bibnumfmt}[1]{[S#1]}
\renewcommand{\citenumfont}[1]{S#1}

\setcounter{equation}{0}
\setcounter{figure}{0}
\setcounter{section}{0}
\setcounter{table}{0}
\setcounter{page}{1}
\makeatletter
\renewcommand{\theequation}{S\arabic{equation}}
\renewcommand{\thefigure}{S\arabic{figure}}
\renewcommand{\thesection}{S-\Roman{section}}
\renewcommand*{\thepage}{S\arabic{page}}
\renewcommand{\bibnumfmt}[1]{[S#1]}
\renewcommand{\citenumfont}[1]{S#1}

\section{Contents}
\begin{enumerate}
    \item \hyperref[sm:scaling]{Scaling arguments for the $A_g$-$\chi$ relation}
    \item \hyperref[sm:singularity]{Crack geometry and validation of the tip square root singularity}
    \item \hyperref[sm:loading]{Constrained uniaxial tension \textit{vs.} constrained tension}
    \item \hyperref[sm:chinu]{$\chi$-$\nu$ and $\mu_{\rm na}/\mu$-$\nu$ relations for MD-quenched- and FSP-glasses}
    \item \hyperref[sm:crackdbt]{Effect of crack length on the brittle-to-ductile transition}
    \item \hyperref[sm:thickness]{Effect of slab thickness on toughness}
    \item \hyperref[sm:toughness]{Effect of initial crack length on glass toughness}
    \item \hyperref[sm:expsim]{Correlations between the abundance of soft defects and the elastic-moduli susceptibility to thermal aging}
\end{enumerate}

\section*{1. Scaling arguments for the $A_g$-$\chi$ relation}
\label{sm:scaling}
In Fig.~3d of the main text we show that for our computer glasses made by quenching liquids at finite cooling rates $\dot{T}$, the quantifiers of mechanical disorder $A_{\rm g}$ and $\chi$ are approximately related by
\begin{equation}
    A_{\rm g} \sim \chi^{10/3}\,.
\end{equation}
Here we explain the origin of this approximate scaling law. In \cite{sm_sticky_spheres_part_1} it was shown that $\chi\!\sim\!\xi^{2/3}$ where $\xi$ is a glassy lengthscale that characterizes responses to local perturbations. In \cite{sm_pinching_pnas} it was demonstrated that $\xi\!\sim\!\omega_{\rm g}^{-1}$, with $\omega_{\rm g}$ being a characteristic frequency scale of soft quasilocalized excitations. Finally, in~\cite{sm_sticky_spheres_part2} it was shown that $A_{\rm g}\!\sim\!\omega_{\rm g}^{-5}$, implying together with the rest of the aforementioned scaling laws that $A_{\rm g}\!\sim\!\chi^{10/3}$, consistent with our data of Fig.~3d of the main text.

\section*{2. Crack geometry and validation of the tip square root singularity}
\label{sm:singularity}
\begin{figure}[ht!]
  \includegraphics[scale=1.15]{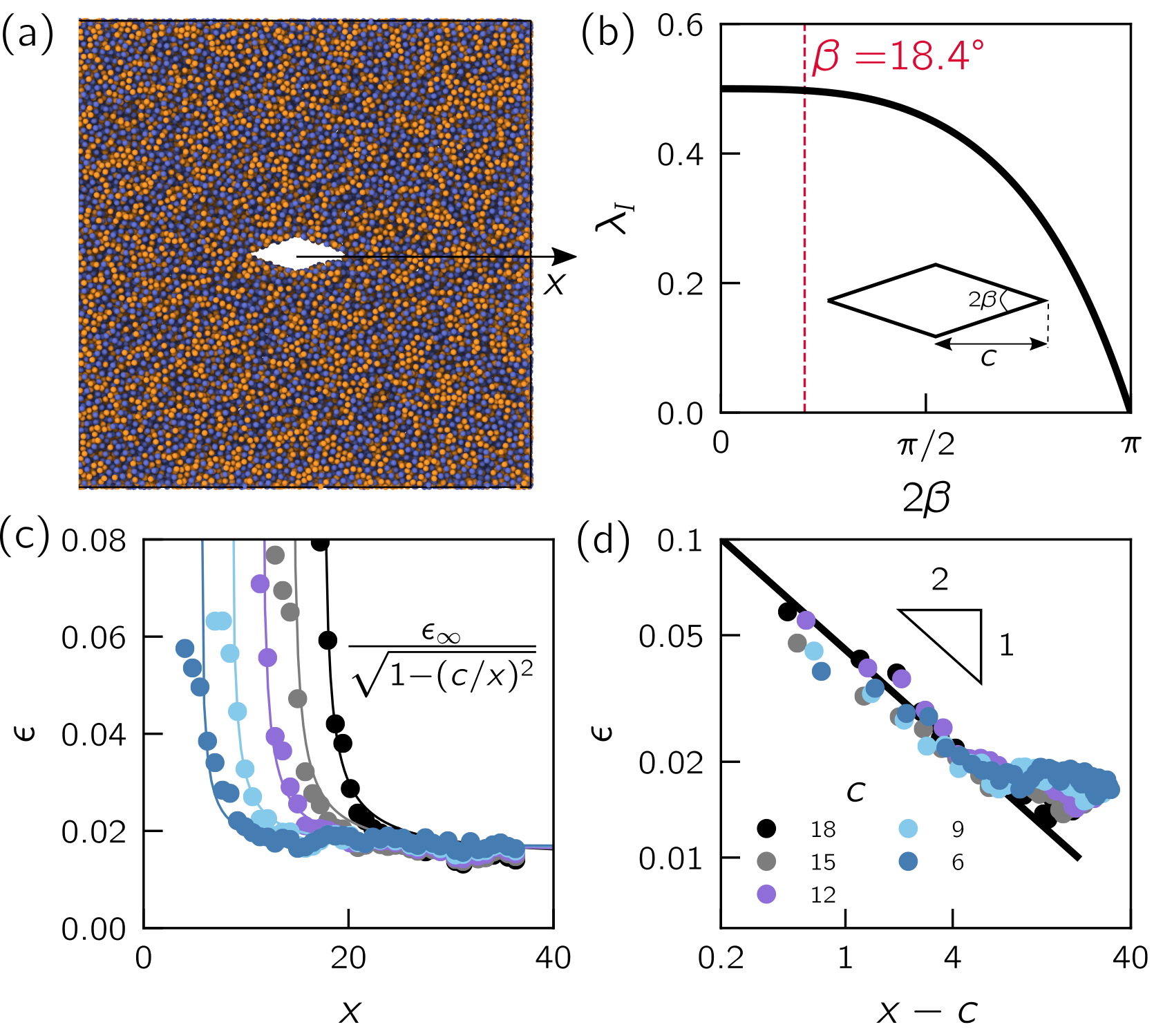}
  \caption{{\bf Crack geometry and square root stress singularity}. (a) Snapshot showing the typical starting configuration of a glass with a diamond shape crack. (b) Dependence of the singularity exponent $\lambda_I$ in mode-I fracture on the vertex angle $2\beta$ (see sketch for definition). The vertical red dashed line indicates the angle used throughout this work. (c) Local extensional strain $\epsilon$ along the $x$-axis, as indicated in (a), for various crack lengths $c$ and fixed global extensional strain $\epsilon_\infty\!=\!0.02$. (d) Same data as in (c), plotted on double logarithmic scales against $x\!-\!c$, demonstrating the square root singularity of the strain close to the notch tip. }
  \label{sm_fig:fig2}
\end{figure}

In Fig.~\ref{sm_fig:fig2}a, we present a 2D projection of the typical starting configuration used throughout our study. The initial crack geometry follows a diamond shape (see sketch in the inset of Fig.~\ref{sm_fig:fig2}b) with semi-axis $c$ and total crack length $2c$. The vertex angle $\beta$ is set to $18.4^{\rm o}$. For such a geometry, the stress singularity at the crack tip will scale with $1/(2\pi r)^{\lambda_I}$, with $r$ being the distance from the crack tip and $\lambda_I$ being the mode-I fracture singularity order~\cite{sm_savruk2010two}. The later depends on the vertex angle $\beta$, and can be approximated by the following form \cite{sm_savruk2010two}
\begin{equation}
\lambda_I(\beta) \simeq 1.247\cos(\beta) - 1.312 \cos^2(\beta)+0.8532 \cos^3(\beta) - 0.2882 \cos^4(\beta)\,,
\end{equation}
which is plotted in Fig.~\ref{sm_fig:fig2}b. It is clear that, for our chosen vertex angle $\beta\!=\!18.4^{\rm o}$ (indicated by the vertical dashed line), $\lambda_I(18.4^{\rm o})\!\approx\!1/2$, and so we can expect to observe the well-known $\sim\!1/\sqrt{r}$ singularity at distance $r$ from the crack tip.

To confirm this expectation, we compute the local strain along the $x$-axis (perpendicular to the loading direction) for a fixed strain of $\epsilon\!=\!2\%$, and different crack lengths $2c$. In practice, the local strain $\epsilon(x)$ is extracted from the $D^2_{\rm min}$ procedure proposed by Falk and Langer in Ref.~\cite{sm_falk_langer_stz}, and the distance $x$ is measured from the center of the crack, as illustrated in Fig.~\ref{sm_fig:fig2}a. For a thin crack, one expects the local strain to take the following form~\cite{sm_rice1968path}
\begin{equation}
\label{eq:staintip}
\epsilon(x)=\frac{\epsilon_\infty}{\sqrt{1-(c/x)^2}},
\end{equation}
with $\epsilon_\infty$ denoting the remote loading strain. Despite the limited size offered by atomistic simulations, we observe in Fig.~\ref{sm_fig:fig2}c a good agreement between the local strain and Eq.~(\ref{eq:staintip}). Shifting the $x$-axis of Fig.~\ref{sm_fig:fig2}c by the semi-crack length $c$, we indeed observe in Fig.~\ref{sm_fig:fig2}d the expected square root singularity of the strain approaching the crack tip.

\section*{3. Constrained uniaxial tension \textit{vs.} constrained tension}
\label{sm:loading}
\begin{figure}[ht!]
  \includegraphics[scale=1.15]{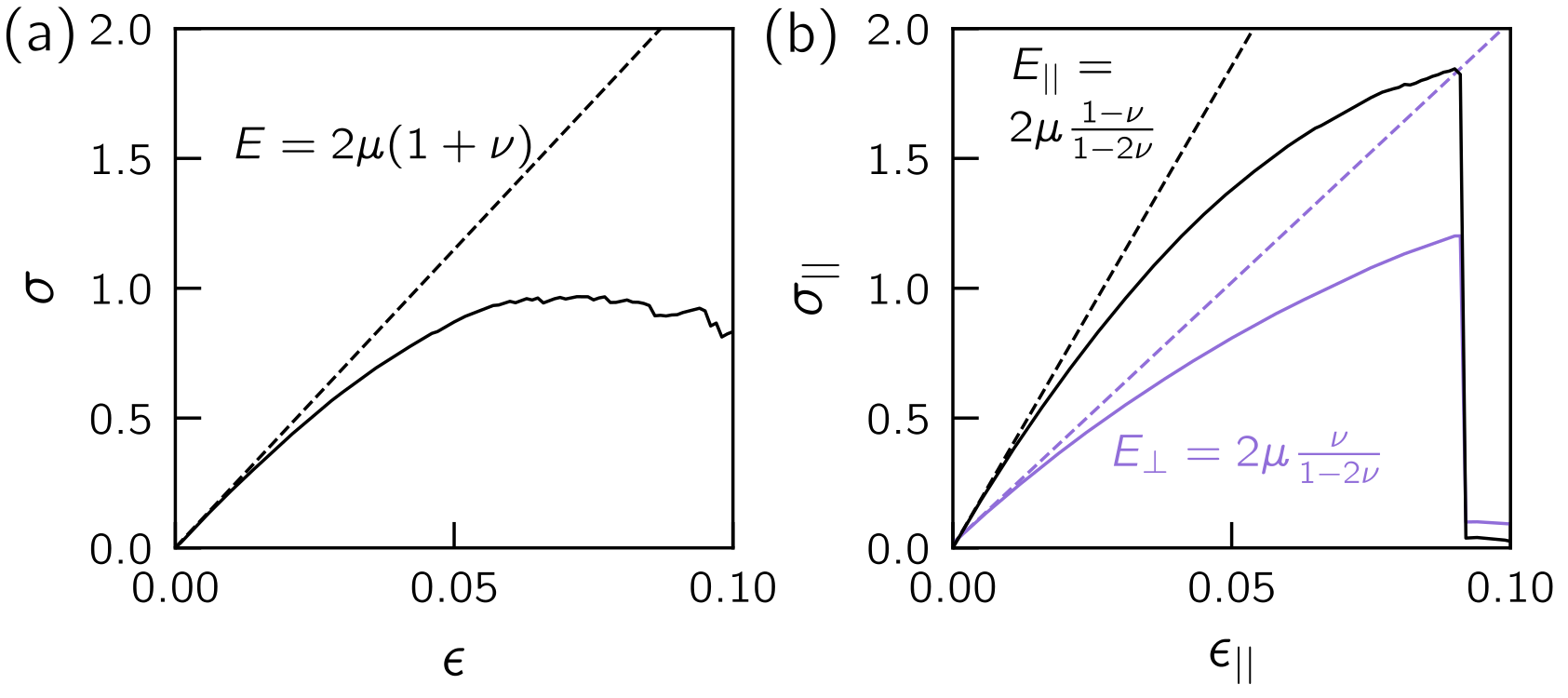}
  \caption{ {\bf Elastic moduli: unconstrained uniaxial tension \textit{vs.}~constrained tension}. (a) We plot a typical stress-strain curve of a glass under unconstrained uniaxial tension. Black dashed line indicates the linear response controlled by the Young modulus. (b) Typical stress-strain curve under constrained tension. The stress parallel and perpendicular to the loading direction are colored in black and purple, respectively. Dashed lines indicate the predicted linear response based on Hooke's law. }
  \label{sm_fig:fig3}
\end{figure}

In Fig.~\ref{sm_fig:fig3}a-b, we validate our expression for the different moduli under constrained uniaxial tension and constrained tension (see \emph{Methods} in the main text). The predicted linear response (dashed lines in Fig.~\ref{sm_fig:fig3}) follows directly from computing independently the athermal shear modulus $\mu$ and the Poisson's ratio $\nu$.

\section*{4. $\chi$-$\nu$ and $\mu_{\rm na}/\mu$-$\nu$ relations for MD-quenched- and FSP-glasses}
\label{sm:chinu}
\begin{figure}[ht!]
  \includegraphics[scale=0.8]{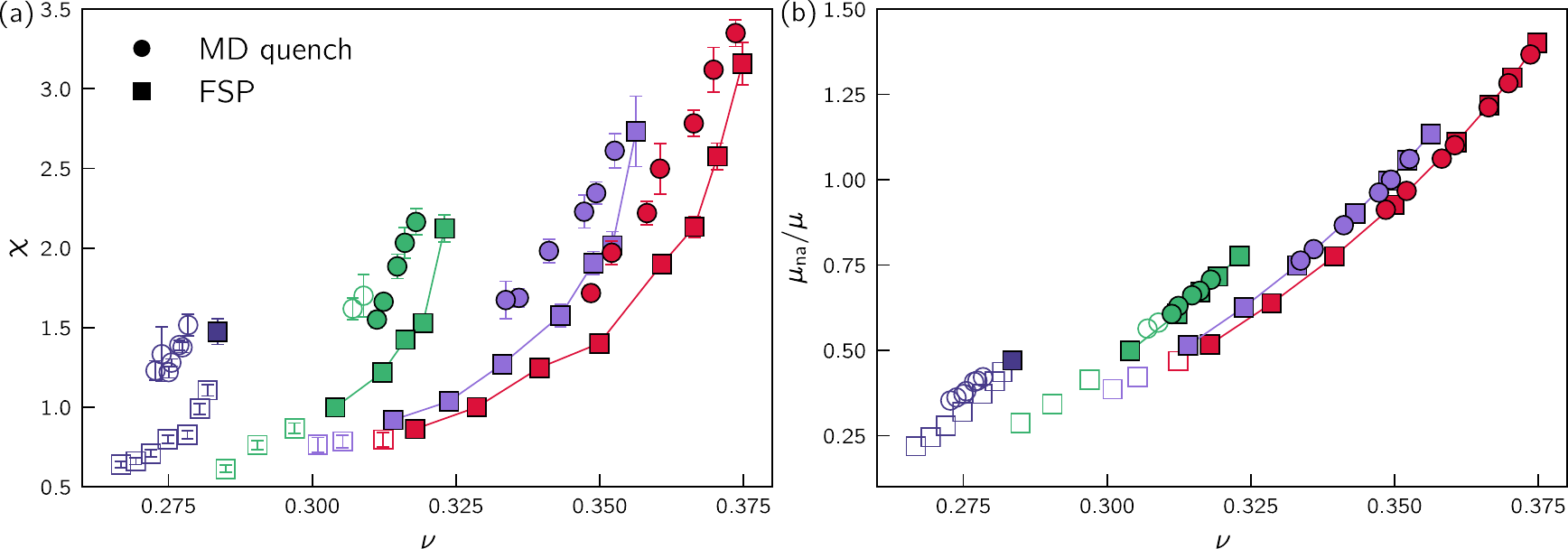}
  \caption{{\bf $\chi$ vs.~$\nu$ and $\mu_{\rm na}/\mu$ vs.~$\nu$ relations.} (a) Disorder quantifier $\chi$ that encapsulates the shear modulus fluctuations, plotted against the Poisson's ratio $\nu$. (b) The ratio $\mu_{\rm na}/\mu$ is plotted against $\nu$. Filled and empty symbols correspond to ductile-like and brittle-like failure, in glass samples with an initial crack length $c\!=\!6$. Circles and squares correspond to MD-quenched and FSP glasses, respectively.}
  \label{sm_fig:fig4}
\end{figure}

In Fig.~\ref{sm_fig:fig4}a, we compare the dependence of the disorder quantifier $\chi$ on the Poisson's ratio $\nu$ between molecular-dynamics (MD) quenches and FSP glasses. We observe that our the $\chi$-$\nu$ curves for FSP glasses and MD-quenched glasses do not fall on the same curves. Instead, we find that $\chi$ measured for FSP glasses is generally lower than that measured for MD-quenched glasses, at the same $\nu$. This discrepancy could be explained by the gapped nonphononic vibrational spectra featured by FSP glasses, as shown in~\cite{fsp}.

We also report in Fig.~\ref{sm_fig:fig4}b the ratio $\mu_{\rm na}/\mu$ between the `non-affine' shear modulus $\mu_{\rm na}$ and the total shear modulus $\mu$, plotted against the Poisson's ratio $\nu$. In contrast to $\chi$, we observe that the $\mu_{\rm na}/\mu$-$\nu$ relation featured by MD-quenched glasses and FSP glasses agree very well, with some small deviation observed for the $r_{\rm c}\!=\!1.2$ glasses. Interestingly, for $r_{\rm c}=1.3$ we observe that some MD-quenched and FSP glasses do not fail in the same way despite having the same $\mu_{\rm na}/\mu$ and $\nu$. We leave for future investigation the use of the SWAP-Monte-Carlo algorithm~\cite{sm_LB_swap_prx} to create deeply aged glasses, which do not exhibit a gapped nonphononic vibrational spectra, in contrast to FSP glasses.

\section*{5. Effect of crack length on the brittle-to-ductile transition}
\label{sm:crackdbt}
The ductile-to-brittle transition presented in Fig.~4 of the main text under unconstrained uniaxial tension pertains to glasses with an initial crack length of $c\!=\!6$. Here we present additional data obtained under variations of the crack length. For MD-quenched glasses, and in the absence of an initial crack, only samples of the $r_{\rm c}\!=\!1.2$ glasses break before 15\% uniaxial strain is reached. Different from our results in the main text for $r_{\rm c}\!=\!1.4$ --- for an initial crack length of $c\!=\!6$ ---, $r_{\rm c}\!=\!1.4$ glasses with an initial crack length of $c\!=\!9$ all exhibit crack propagation.

We perform the same analysis for FSP glasses. We find that without the presence of an initial crack, we observe cavitation for all $r_{\rm c}$'s, but only for glasses featuring extremely low disorder, namely $\chi\!<\!0.8$. As for MD-quenched glasses, the ductile-to-brittle transition moves toward higher $\chi$ values when increasing the crack length. This trend highlights the effect of stress focusing at the crack tip, facilitating void nucleation and subsequently crack propagation.

\begin{figure*}[ht!]
  \includegraphics[scale=0.9]{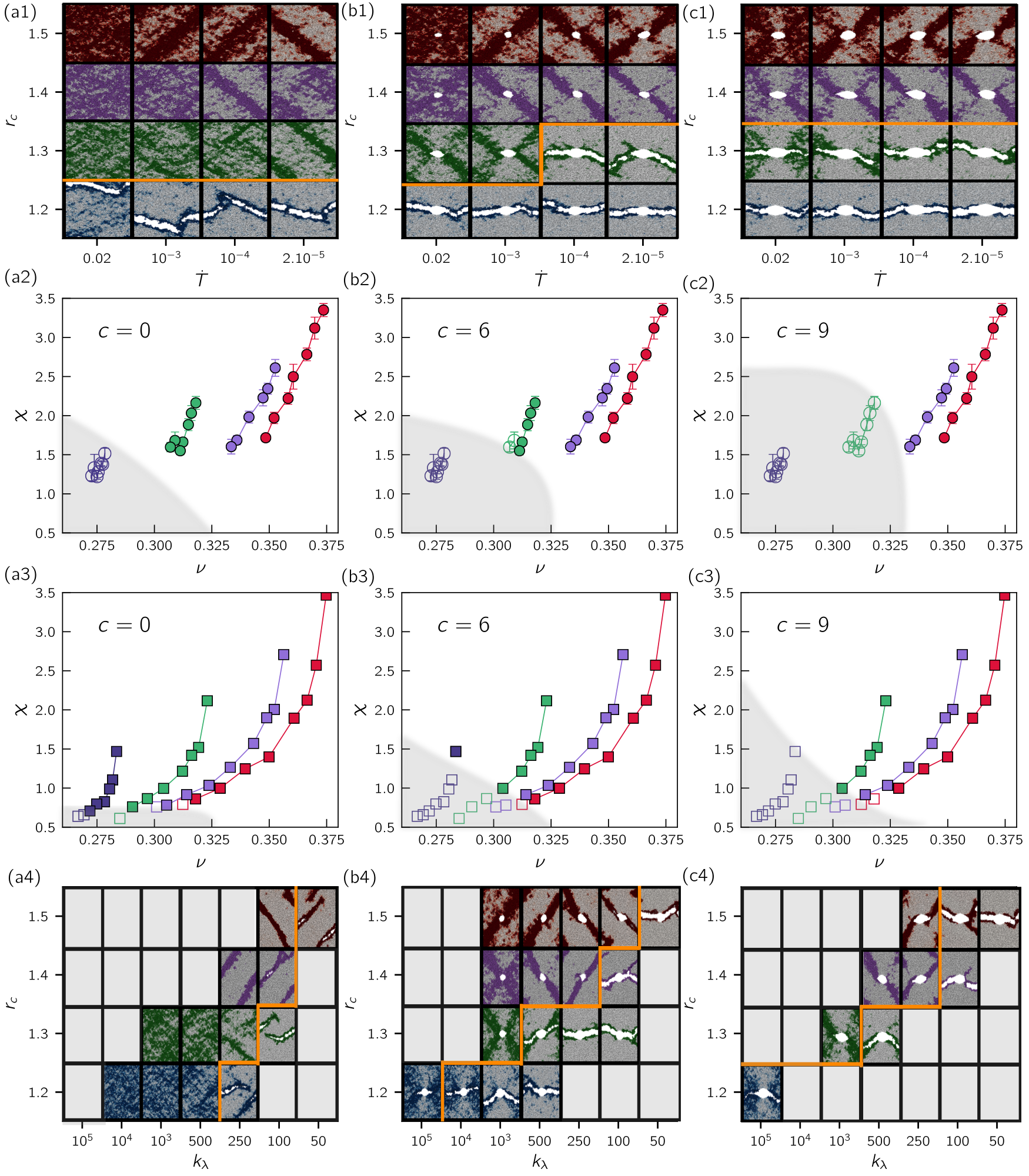}
  \caption{ {\bf Effect of crack length $c$ on the ductile-to-brittle transition}. (a1-c1) Snapshots of the non-affine plastic deformation at $15\%$ uniaxial strain for MD-quenched glasses in the $r_{\rm c}\!-\!\dot{T}$ plane, for crack lengths $c\!=\!0$ (a1), $c\!=\!6$ (b1), and $c\!=\!9$ (c1). (a2-c2) Ductile-brittle phase diagram in the $\chi\!-\!\nu$ plane for MD-quenched glasses and crack lengths $c\!=\!0$ (a2), $c\!=\!6$ (b2), and $c\!=\!9$ (c2). Glasses featuring ductile-like (brittle-like) failure are represented by full (empty) symbols. Blurry regions are added only as guides to the eye. (a3-c3) Same as (a2-c2), for FSP glasses. (a4-c4) Snapshots of the non-affine plastic deformation at $15\%$ strain for FSP glasses in the $r_{\rm c}\!-\!k_\lambda$ plane, for $c\!=\!0$ (a4), $c\!=\!6$ (b4), and $c\!=\!9$ (c4).}
  \label{sm_fig:fig5}
\end{figure*}

\cleardoublepage
\section*{6. Effect of slab thickness on toughness}
\label{sm:thickness}
\begin{figure}[ht!]
  \includegraphics[scale=1.15]{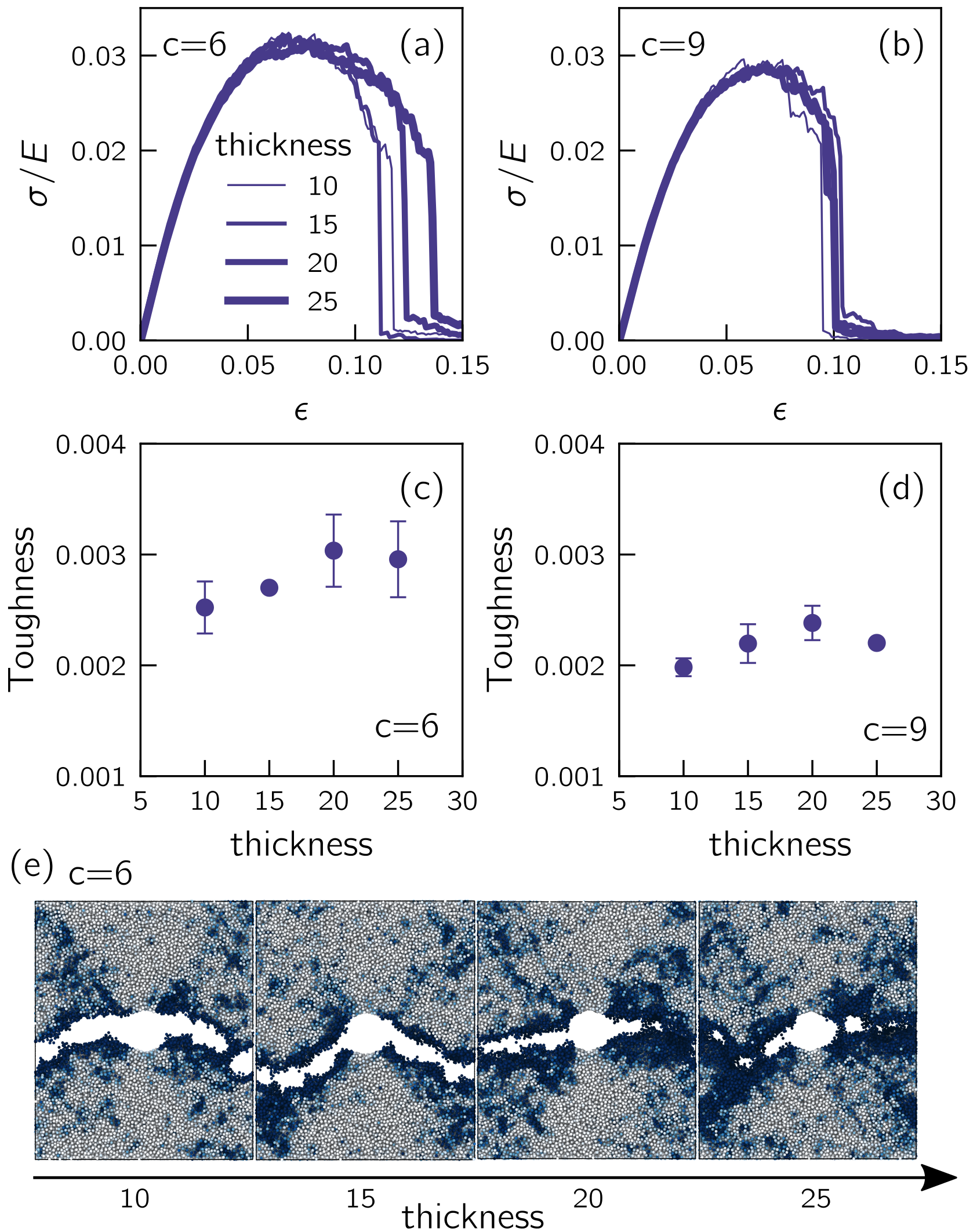}
  \caption{{\bf Effect of sample thickness on the toughness under unconstrained uniaxial tension}. (a) Stress-strain curves under unconstrained uniaxial tension for FSP glasses ($r_{\rm c}=1.2$ and $k_\lambda=1000$) with various slab thicknesses and a fixed initial crack length $c=6$. (b) Same data as in (a) but for an initial crack length of $c=9$. (c) Toughness as a function of the sample thickness for a fixed initial crack length $c\!=\!6$. (d) Same data as in (c) but for a crack length of $c\!=\!9$. (e) Snapshots of the non-affine plastic deformation at $15\%$ uniaxial strain for samples with different thicknesses and a fixed initial crack of $c\!=\!6$.}
  \label{sm_fig:fig6}
\end{figure}

Throughout our study presented in the main text, the sample thickness was fixed to $15$ particles diameters. It is well known that the fracture toughness of a material does depend on the sample thickness. A complete study of the effect of sample thickness on the toughness and failure mode is beyond the scope of our present manuscript. We nevertheless report here the effect of sample thickness on the toughness under unconstrained uniaxial test for one glass realization: an FSP-glass prepared with $k_\lambda\!=\!1000$ and potential cutoff $r_{\rm c}=1.2$. We find a noticeable increase of the toughness for a small crack length ($c\!=\!6$, see Fig.~\ref{sm_fig:fig6}a and Fig.~\ref{sm_fig:fig6}c), whereas we observe for thicker samples an enhanced necking during fracture (see Fig.~\ref{sm_fig:fig6}e). For a larger crack length, namely $c=9$, we find the toughness to be nearly independent of the sample thickness (see Fig.~\ref{sm_fig:fig6}b and Fig.~\ref{sm_fig:fig6}d).

\cleardoublepage

\section*{7. Effect of initial crack length on glass toughness}
\label{sm:toughness}
\begin{figure*}[ht!]
  \includegraphics[scale=1]{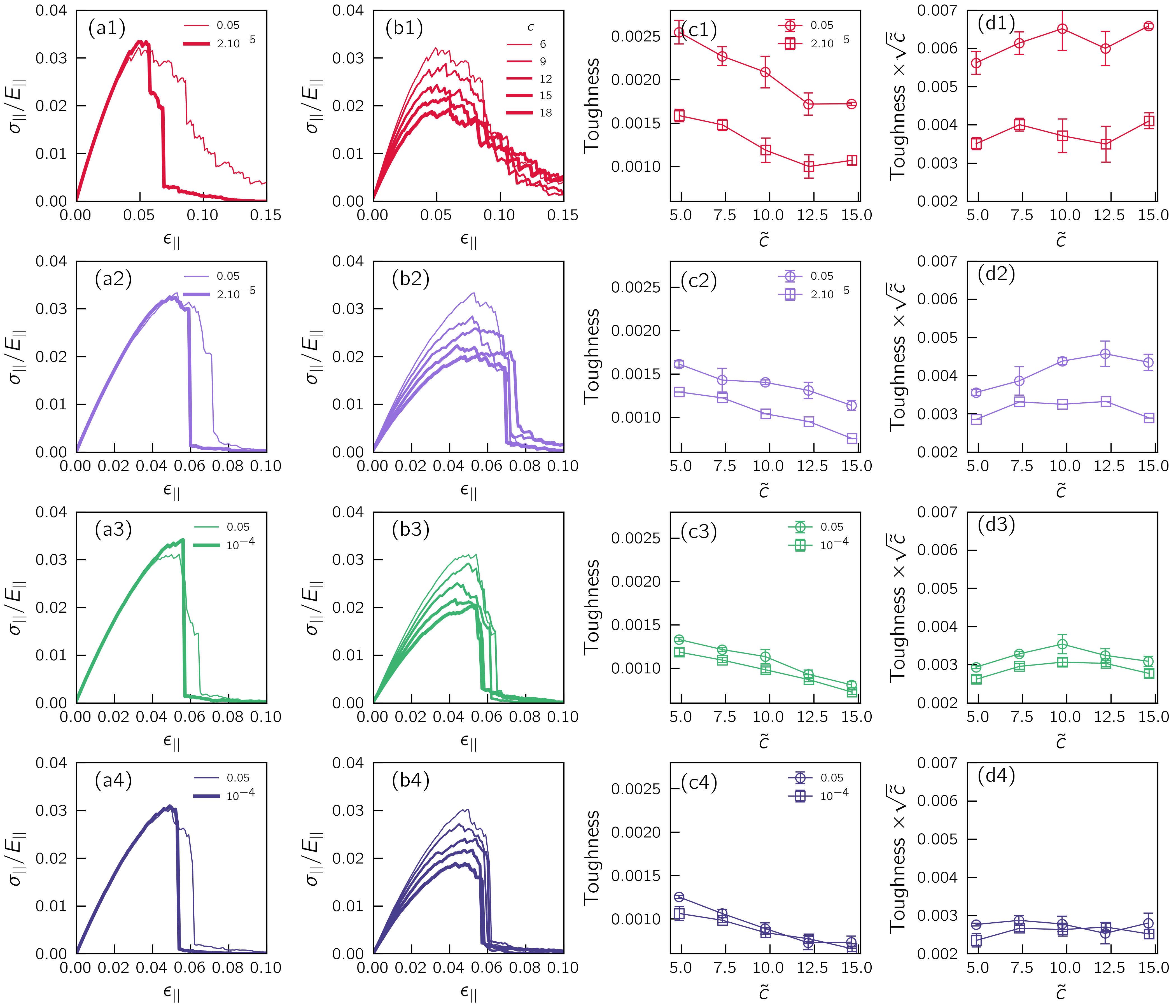}
  \caption{{\bf Effect of initial crack length on glass toughness under constrained tension}. (a) Stress-strain curves under constrained tension, for glasses prepared at a high quench rate (thin curve) and low quench rate (thick curve). See text and \emph{Methods} for the exact definitions of $\sigma_\|$, $\epsilon_\|$ and $E_\|$. (b) The same as in panel (a), but for for glasses prepared at a high quench rate $\dot{T}\!=\!0.02$ and different crack length $c$. (c) The Toughness (computed as the integral over the stress-strain curve, see text for details, and normalized by $E_\|$) as a function of the normalized crack length $\tilde{c}=c/a_0$ (with $a_0=\rho^{-1/3}$) for glasses prepare at a high (circles) and low (squares) quench rates. (d) Same data as in panel (c) but where the toughness is scaled by the square root of the crack length. Error-bars are estimated from sample-to-sample fluctuations. From the top to the bottom row are data for $r_{\rm c}\!=\!1.5$, $1.4$, $1.3$, and $1.2$, respectively.}
  \label{sm_fig:fig7}
\end{figure*}

The results for glass toughness under constrained tension --- presented in Fig.~6 of the main text --- were only reported for an initial crack length with $c\!=\!6$. Here we provide additional data pertaining to various initial crack lengths. Figure~\ref{sm_fig:fig7} is organized as follows; we first show in columns ($a_x$) typical stress-strain curves for poorly and well-aged MD-quenched glasses. Columns ($b_x$) show the influence of the crack length on the stress-strain curves for a fixed quench rate (here $\dot{T}\!=\!0.02$). In columns ($c_x$) we report the dimensionless toughness as a function of the dimensionless crack length $\tilde{c}\!=\!c/a_0$ ($a_0$ is an interparticle distance) for poorly (circles) and well-aged (squares) glasses. Finally, we show that the dimensionless toughness measure times the square root of the crack length $\sqrt{\tilde{c}}$ is largely independent of the crack size, and thus controlled by the square root singularity of the stress at the tip.

\cleardoublepage

\section*{8. Correlations between the abundance of soft defects and the elastic-moduli susceptibility to thermal aging}
\label{sm:expsim}
In this final Section we discuss the relation between macroscopic and microscopic materials properties, and the susceptibility of a glass's shear modulus to thermal annealing. In particular, we consider the number density of soft spots --- as embodied in the prefactor $A_{\rm g}$ --- and the Poisson's ratio $\nu$, in as-cast, highly disordered glasses. As explained in the \emph{Methods} section of the main text, the athermal shear modulus can be decomposed into an affine term $\mu_{\rm af}$ (Born term) and a non-affine term $\mu_{\rm na}$ as $\mu\!=\!\mu_{\rm af}-\mu_{\rm na}$. The total modulus reads
\begin{equation}
    \mu \equiv \frac{1}{V}\left(\frac{\partial^2U}{\partial\gamma^2} - \frac{\partial^2U}{\partial\gamma\partial\xv}\cdot\calBold{H}^{-1}\cdot\frac{\partial^2U}{\partial\xv\partial\gamma} \right)\,
\end{equation}
where $\calBold{H}\!\equiv\!\frac{\partial^2U}{\partial\xv\partial\xv}$ is the Hessian matrix of the potential energy $U$, which captures the physics of soft excitations (plasticity carriers) in the system. In the case of highly disordered samples, $\calBold{H}$ embodies a high density of soft excitations that are known to couple with the shear-induced forces $\frac{\partial^2U}{\partial\xv\partial\gamma}$. The effect of soft modes is to increase $\mu_{\rm na}$ and decrease $\mu$, which ultimately decreases the shear to bulk moduli ratio $\mu/K$ and therefore increases the Poisson's ratio $\nu$. Overall, we thus expect that samples with a high $A_{\rm g}$ will feature a large Poisson's ratio.

\begin{figure*}[ht!]
  \includegraphics[scale=1]{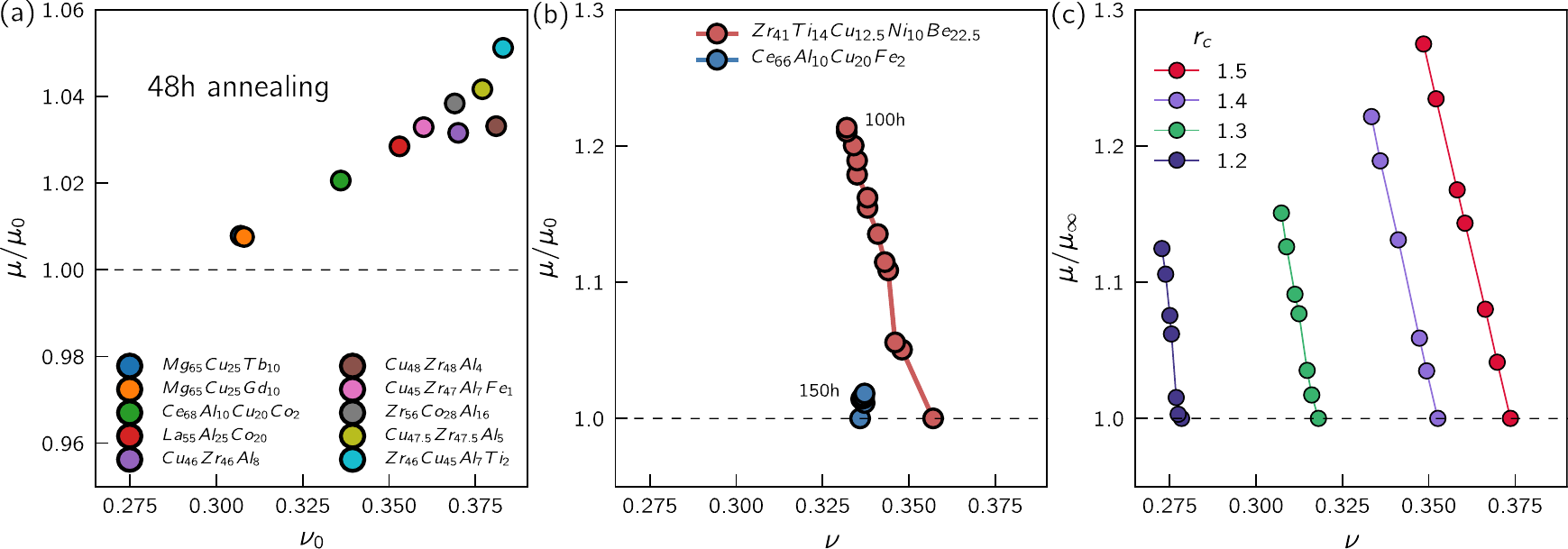}
  \caption{{\bf Effect of thermal aging on macroscopic elasticity: laboratory bulk metallic glasses \textit{vs.}~computer glasses}. (a) Correlation between the aging-induced stiffening of the shear modulus $\mu/\mu_0$ and as-cast Poisson's ratio $\nu_0$ for various bulk metallic glasses (BMGs) annealed for 48 hours (taken from Ref.~\cite{sm_wang2014understanding}). (b) Rescaled shear modulus $\mu/\mu_0$ plotted against $\nu$ for two BMGs, showing either a weak or substantial aging-induced stiffening (taken from Ref.~\cite{sm_sun2016correlation}). (c) Same as in (b) but for our MD-quenched glasses with various $r_{\rm c}$, where the shear modulus is normalized by the high-quench-rate modulus $\mu_\infty$.}
  \label{sm_fig:fig1}
\end{figure*}

Upon thermal aging, $A_{\rm g}$ decreases (up to one order of magnitude for $r_{\rm c}=1.5$, and a factor $2-3$ for $r_{\rm c}=1.2$), which in turn lowers $\mu_{\rm na}$ and increases the total shear modulus. Therefore, glasses with the largest variation in $A_{\rm g}$ are excepted to also show the largest stiffening of $\mu$. Finally, we also expect that as-cast glasses with the largest Poisson's ratio will exhibit a large variability in their shear modulus upon thermal aging. These trends are demonstrated in Fig.~\ref{sm_fig:fig1}a-b with experimental data for bulk metallic glasses (digitized from Refs.~\cite{sm_wang2014understanding,sm_sun2016correlation}), and in Fig.~\ref{sm_fig:fig1}c for our computer glasses with various interaction parameters $r_{\rm c}$ as indicated by the legend.

%

\end{document}